\documentclass[prb,twocolumn,showpacs,preprintnumbers,amsmath,amssymb,floatfix]{revtex4}

\usepackage{graphicx,color}

\begin{document}

\title{Quantum Monte Carlo calculation of the zero-temperature phase diagram
  of the two-component fermionic hard-core gas in two dimensions}

\author{N.\ D.\ Drummond}

\affiliation{TCM Group, Cavendish Laboratory, University of Cambridge, J.\ J.\
Thomson Avenue, Cambridge CB3 0HE, United Kingdom and Department of Physics,
Lancaster University, Lancaster LA1 4YB, United Kingdom}

\author{N.\ R.\ Cooper}

\affiliation{TCM Group, Cavendish Laboratory, University of Cambridge, J.\ J.\
Thomson Avenue, Cambridge CB3 0HE, United Kingdom}

\author{G.\ V.\ Shlyapnikov}

\affiliation{Laboratoire de Physique Th\'eorique et Mod\`eles Statistiques,
Universit\'e Paris Sud, CNRS, 91405 Orsay, France}

\affiliation{van der Waals-Zeeman Institute, University of Amsterdam,
Valckenierstraat 65/67, 1018 XE Amsterdam, The Netherlands}

\author{R.\ J.\ Needs}

\affiliation{TCM Group, Cavendish Laboratory, University of Cambridge, J.\ J.\
Thomson Avenue, Cambridge CB3 0HE, United Kingdom}

\date{\today}

\begin{abstract}
Motivated by potential realizations in cold-atom or cold-molecule systems, we
have performed quantum Monte Carlo simulations of two-component gases of
fermions in two dimensions with hard-core interactions. We have determined the
gross features of the zero-temperature phase diagram, by investigating the
relative stabilities of paramagnetic and ferromagnetic fluids and crystals.
We have also examined the effect of including a pairwise, long-range $r^{-3}$
potential between the particles.  Our most important conclusion is that there
is no region of stability for a ferromagnetic fluid phase, even if the
long-range interaction is present.  We also present results for the
pair-correlation function, static structure factor, and momentum density of
two-dimensional hard-core fluids.
\end{abstract}

\pacs{67.85.Lm, 02.70.Ss, 67.10.Db}

\maketitle

\section{Introduction \label{sec:introduction}}

Ultracold gases of fermionic atoms and molecules have been the subject of a
large number of experimental and theoretical studies\cite{blochdz,giorgini} in
the last decade.  The interest in these systems arises from the fact that the
atoms or molecules obey quantum statistics rather than classical statistics
and exhibit interesting and novel quantum phases at sufficiently low
temperatures.  Unlike electron systems, it is possible to manipulate the
interaction between the atoms or molecules to some extent, e.g., by Feshbach
resonances\cite{chin_2010} and/or by applying microwave
fields.\cite{papoular,innsbruck,gorshkov_2008,cooper_2009} The ability to
control the interactions can be used to gain insight into the physics of phase
transitions and the nature of correlation effects in these quantum phases.
Celebrated examples are the experimental studies of the Mott transition for
bosons\cite{greiner} and of the superfluid pairing in the unitary
(strong-coupling) Fermi gas.\cite{blochdz,giorgini} Finally, ultracold atomic
gases may play an important role in future quantum computing devices.

Recently, interest has turned towards the use of ultracold atomic systems to
investigate ferromagnetism in Fermi gases.  Itinerant ferromagnetism in
electron systems is poorly understood, and it is possible that insights might
be gained by studying ferromagnetic fluid states in cold-atom
systems.\cite{duine} Experimental studies of strongly repulsive Fermi gases
close to Feshbach resonances have found behavior consistent with the formation
of (local) ferromagnetic correlations in the manner of Stoner
ferromagnetism.\cite{ketterle} However, these systems are limited by an
intrinsic instability to molecule formation, which results in a short lifetime
for the ferromagnetic state and may also complicate the interpretation of
experimental features.\cite{pekker,babadi} It is therefore of interest to
explore other forms of interaction that might give rise to ferromagnetism in
the absence of nearby bound states.

An interesting class of experimental system in this regard is provided by
gases of fermionic polar molecules (e.g., $^{40}$K$^{87}$Rb or
$^7$Li$^{40}$K). Polar molecules that are confined to two-dimensional (2D)
layers and dressed by a circularly polarized microwave field experience a
long-range potential that falls off as $r^{-3}$, where $r$ is the
intermolecular distance.\cite{innsbruck,gorshkov_2008,cooper_2009}
Furthermore, the dressed states have the feature that there is a very sharp
increase in the potential energy at short distances, preventing close
approach.  Hence such molecules can be modeled as a gas of fermionic hard-core
particles with an additional long-range potential varying as $r^{-3}$. The
hard-core potential prevents the formation of shallow bound states even for
(weakly) attractive long-range interactions. It has been suggested that a
single-component gas of such fermionic molecules will exhibit a topological
superfluid $p$-wave pairing phase, of direct relevance to possible
topologically protected quantum computing
devices.\cite{cooper_2009,nayak_2008}

Here we consider the case of a two-component gas of such fermionic molecules.
We have applied quantum Monte Carlo (QMC) methods to address the issue of
whether or not the hard-core interactions cause this system to exhibit a
region of itinerant ferromagnetism.  In particular we have used the
variational and diffusion quantum Monte Carlo (VMC and DMC)
techniques\cite{ceperley_1980,foulkes_2001} to establish the ground-state
phase diagram of two-component gases of fermionic hard-core particles moving
in 2D\@.  QMC methods are widely acknowledged to be the most accurate
first-principles techniques available for studying condensed matter.  In the
VMC method, Monte Carlo integration is used to evaluate expectation values
with respect to an explicitly correlated many-body trial wave function.  In
the DMC method, the ground-state component of a trial wave function is
projected out by a stochastic algorithm.  Fermionic symmetry is maintained by
the fixed-node approximation.\cite{anderson_1976}

Some of the earliest applications of QMC methods were to study the
ground-state properties of bosonic hard-sphere fluids, as a model for the
behavior of $^4$He.\cite{liu_1973,demichelis_1978,lei_1990} More recently, QMC
methods have been used to study 3D fermionic systems in which there is a
hard-sphere repulsion between opposite-spin
particles.\cite{chang_2010,pilati_2010} To the best of our knowledge, however,
QMC methods have never previously been used to study the phase diagram of 2D
gases of fermionic hard-core particles

This paper is arranged as follows.  In Sec.\ \ref{sec:model} we describe the
model Hamiltonian that we study.  In Sec.\ \ref{sec:wave_functions} we discuss
the form of trial wave function that we use.  In Sec.\ \ref{sec:finite_size}
we consider the issue of finite-size errors and explain how we extrapolate to
the thermodynamic limit.  In Sec.\ \ref{sec:phase_diagram} we present our
results for the phase diagram of fermionic hard-core particles and in Sec.\
\ref{sec:expvals} we present results for the pair-correlation function (PCF),
static structure factor, and momentum density of hard-core fluids.  In Sec.\
\ref{sec:attractive_interactions} we investigate whether the inclusion of
weak, long-range interactions affects the phase diagram.  Finally, we draw
our conclusions in Sec.\ \ref{sec:conclusions}.  All our QMC calculations were
performed using the \textsc{casino} code.\cite{casino}

\section{Hard-core model \label{sec:model}}

\subsection{Hard-core Hamiltonian}

Suppose we have a system of interacting, circular hard-core particles moving
in two dimensions.  Throughout, we use units (a.u.)\ in which the Dirac
constant, the mass of the particles, and the radius of the circle that
contains one particle on average are unity.  We assume the particles to be
fermions with spin $s=1/2$.  Let $D$ be the diameter of the particles.  In
these units the Hamiltonian is
\begin{equation} \hat{H} = \sum_i - \frac{1}{2}
\nabla_i^2 + \sum_{i>j} v_H(r_{ij}), \end{equation} where
\begin{equation} v_H(r)=\left\{ \begin{array}{ll} 0
& {\rm if}~r>D \\ \infty & {\rm otherwise} \end{array} \right. .
\end{equation}
When we refer to ``high density,'' we mean that the value of $D$ is large
(comparable with the radius of the circle that contains one particle on
average).  In our calculations we used finite numbers $N$ of particles in
cells subject to periodic boundary conditions.  From these we extrapolated to
infinite system size, as discussed in Sec.\ \ref{sec:finite_size}.

The potential energy is zero throughout the permitted region of configuration
space. Hence the expected potential energy is zero, and the system only has
kinetic energy.  Nevertheless, the hard-core interaction has a significant
effect on the energy, since it defines the boundary conditions on the solution
of the Schr\"{o}dinger equation.  Hence the system does not behave like a
noninteracting Fermi gas: the magnetic behavior is nontrivial and the system
must crystallize at a sufficiently high density.

Hard-core systems have a maximum density: the close-packing limit.  In 2D the
triangular lattice has the highest packing fraction, with the maximum
hard-core diameter being $D_{\rm max} = \sqrt{2\pi/\sqrt{3}}$.

\subsection{Qualitative features of the phase diagram}

For infinitesimal $D$, the system resembles a noninteracting Fermi gas.  The
momentum density is a unit step function.  The paramagnetic fluid phase is
favored, because it has the lowest kinetic energy.  The crystal is not even
stable as an excited state for small $D$.

As $D$ is increased, the energy of the paramagnetic fluid rises more rapidly
than that of the ferromagnetic fluid, because wave-function antisymmetry
already keeps parallel-spin particles apart. The hard-core interactions have a
greater effect on the distribution of antiparallel-spin particles, with the
short-range PCF being forced to go to zero, as shown in Sec.\ \ref{sec:pcf}.
The momentum density of a hard-core fluid with $D>0$ is not a unit step
function, but it retains a discontinuity at the Fermi wave vector $k_F$.  The
crystal becomes stable as $D$ is increased, but with a higher energy than that
of the fluid.

At large $D$, the energies of the ferromagnetic and (frustrated)
antiferromagnetic crystals are very similar.  At some value of $D$ the
ferromagnetic fluid is lower in energy than the paramagnetic fluid, but at
some other value of $D$ the crystal becomes more stable than either of the
fluid phases.  It is a nontrivial problem to determine which of these
transitions occurs first.  The crystal orbitals become delta functions in the
close-packing limit.

\section{Trial wave functions \label{sec:wave_functions}}

\subsection{Slater-Jastrow-backflow wave functions}

We used trial wave functions of Slater-Jastrow-backflow form
$\Psi=\exp(J)S^\uparrow S^\downarrow$.  The Jastrow exponent $J$ included
polynomial and plane-wave functions of the interparticle
distances\cite{ndd_jastrow} together with a two-body term to impose the
hard-core boundary conditions, as described in Sec.\ \ref{sec:hard_core_bcs},
and a three-body term.\cite{casino_manual} The backflow function consisted of
a polynomial expansion in the interparticle distance.\cite{backflow}  The
terms in the Jastrow exponent and backflow function arising from parallel-spin
and antiparallel-spin pairs of particles were allowed to differ.

Plane-wave orbitals $\exp(i{\bf k} \cdot {\bf r})$ were used in the Slater
determinants $S^\uparrow$ and $S^\downarrow$ for the fluid phases.  The fluid
phases suffer from momentum quantization effects (single-particle finite-size
errors).  To reduce these, we performed twist averaging in the canonical
ensemble.\cite{lin_twist_av}

We used Gaussian orbitals $\exp \left( -C|{\bf r}-{\bf R}_p|^2 \right)$
centered on hexagonal lattice sites $\{{\bf R}_p\}$, where the exponent $C$ in
the crystal orbitals is an optimizable parameter, for the crystal phases.

These calculations are similar to the QMC calculations that have been
performed to establish the zero-temperature phase diagram of the three
dimensional homogeneous electron gas
(HEG)\cite{ceperley_1980,zong_2002,ndd_3d_heg} and 2D
HEG\@.\cite{tanatar,rapisarda,ndd_2d_heg}

\subsection{Hard-core two-body behavior \label{sec:hard_core_bcs}}

\subsubsection{Antiparallel spins \label{sec:antiparallel_bcs}}

Let us rewrite the Schr\"{o}dinger equation for two hard-core particles in
terms of the center-of-mass and difference coordinates.  We assume the center
of mass is in its zero-energy ground state.  The Schr\"{o}dinger equation for
the difference coordinate ${\bf r}$ is
\begin{equation} -\nabla^2 \Psi = E \Psi. \end{equation}
For the boundary conditions we assume that $\Psi=0$ at $r=D$ and $\Psi=0$ at
$r=R$, where $R$ is very large.  By increasing $R$, the ground-state
eigenvalue $E$ can be made arbitrarily close to 0.  $\Psi$ is circularly
symmetric in the ground state for antiparallel-spin particles, so
\begin{equation} -\frac{\partial^2 \Psi}{\partial r^2} -
\frac{1}{r} \frac{\partial \Psi}{\partial r} \approx 0,
\end{equation}
with general solution $\Psi \approx A+B \log(r)$.  This approximation is valid
over any range of $r$ that is small compared with $R-D$.  However, we are
interested in particle separations of the order $D$ and slightly larger.
Applying the boundary condition $\Psi(D)=0$ gives
\begin{equation} \Psi(r) \propto
    \log(r/D), \end{equation} suggesting the antiparallel-spin two-body
Jastrow factor in a hard-core gas should be $\log(r/D)$.

The two-body behavior of hard-core gases was studied in Ref.\ \onlinecite{li},
but this work did not give the two-body Jastrow factor for the 2D gas and did
not consider parallel-spin pairs.

\subsubsection{Parallel spins \label{sec:parallel_bcs}}

Suppose the two hard-core particles have parallel spins.  In the lowest-energy
state the difference-coordinate wave function is of the form
$\Psi=W(r)r\cos(\theta)$ and the energy eigenvalue is zero in the limit that
the region over which the wave function is normalized is large.  Hence the
Schr\"{o}dinger equation for the radial part is
\begin{equation} - \frac{d^2W}{dr^2}-\frac{3}{r} \frac{dW}{dr}
\approx 0. \end{equation} The general solution over a range of $r$ values that
is small compared with the region over which the wave function is normalized
is $W(r) \approx A+Br^{-2}$. Applying the boundary condition $W(D)=0$, one
obtains $W(r) \propto 1-D^2/r^2$.  So, for small $r$,
\begin{equation} \Psi \propto \left( 1- \frac{D^2}{r^2} \right) r
\cos(\theta), \end{equation} suggesting the parallel-spin two-body Jastrow
factor in a hard-core gas should be $1-D^2/r^2$.

\subsection{Hard-core Jastrow factor}

The wave function of a hard-core system must go linearly to zero as the
separation of any pair of particles approaches $D$.  To impose this behavior,
the following term was included between all pairs of particles in the Jastrow
exponent $J$ (in addition to the polynomial and plane-wave terms):
\begin{equation} u_H(r)= \left\{ \begin{array}{ll} -\infty & {\rm
if}~r\leq D \\ \log \left[ \tanh \left( \frac{r/D-1}{\alpha(1-r/L_{\rm WS})}
\right) \right] & {\rm if}~D<r<L_{\rm WS} \\ 0 & {\rm if}~r \geq L_{\rm WS}
\end{array} \right. , \end{equation} where $L_{\rm WS}$ is the
radius of the circle inscribed in the Wigner-Seitz cell of the simulation
cell. The parameter $\alpha$ was fixed at 1.  $u_H$ goes smoothly to zero at
$L_{\rm WS}$.  The other terms in the Jastrow factor are analytic at $r=D$.

\begin{figure}
\begin{center}
\includegraphics[clip,scale=0.3]{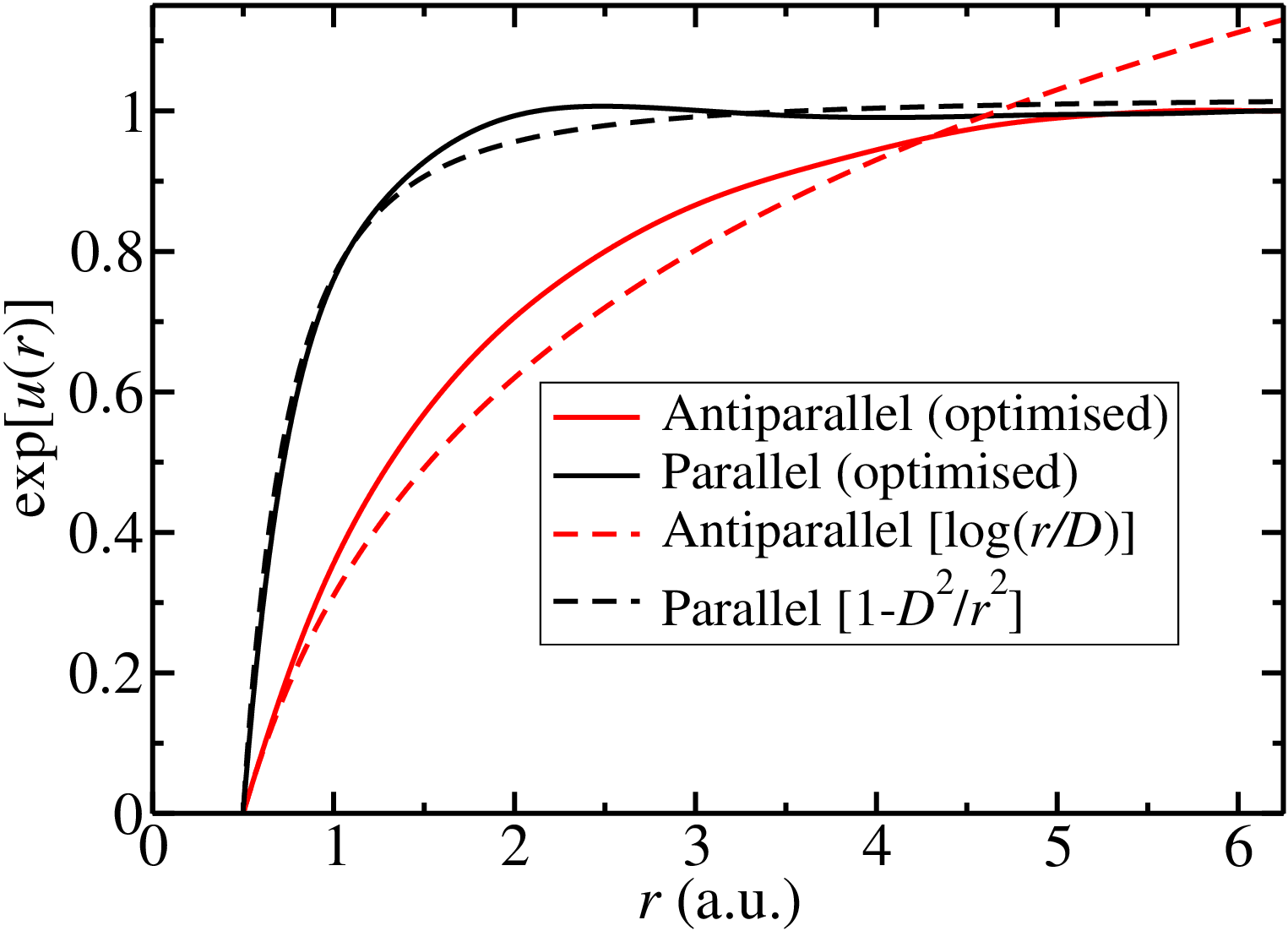}
\caption{(Color online) Two-body Jastrow factors $\exp[u(r)]$ for $N=50$
paramagnetic hard-core particles with $D=0.5$ a.u. \label{fig:jastrow_u}}
\end{center}
\end{figure}

$\exp(u_H) \approx \tanh[(r-D)/D]$ is not of the form $\log(r/D)$ or
$1-D^2/r^2$ suggested by the analytic results for opposite-spin or same-spin
hard-core particles, respectively, but the short-range behavior is correct
[linear in $(r-D)$ in the vicinity of $r=D$]. When optimized, the polynomial
and plane-wave terms in the Jastrow exponent describe all two-body
correlations.  Optimized two-body Jastrow factors are plotted in Fig.\
\ref{fig:jastrow_u}, confirming that the simple theory of two-body
correlations described in Sec.\ \ref{sec:hard_core_bcs} is approximately valid.

\subsection{Free-particle limit}

It is interesting to consider the approach to the free-particle limit as the
diameter $D$ tends to zero.  In a three-dimensional hard-core gas, similar
arguments to those given in Secs.\ \ref{sec:antiparallel_bcs} and
\ref{sec:parallel_bcs} show that the two-body Jastrow factors are
approximately $1-D/r$ and $1-D^3/r^3$ for antiparallel and parallel spins,
respectively, both of which tend to unity as $D \rightarrow 0$.  Hence a
Slater-Jastrow wave function for a three-dimensional hard-core fluid reduces
to a Slater determinant of plane-wave orbitals in the limit of zero particle
diameter.  In the 2D hard-core gas the two-body Jastrow factors are
approximately $\log(r/D)$ and $1-D^2/r^2$ for antiparallel and parallel spins,
respectively.  The antiparallel-spin two-body Jastrow factor is the marginal
case in which two-body correlations become negligible over any given length
scale as $D$ is made small so that free-particle behavior is recovered in the
$D\rightarrow 0$ limit.  In one dimension, however, the infinite contact
potential that remains when the $D\rightarrow 0$ limit is taken prevents the
hard-core gas from exhibiting free-particle behavior, unless the system is
fully ferromagnetic.

\subsection{Behavior of the local energy as hard-core particles collide}

Suppose two antiparallel-spin hard-core particles 1 and 2 approach each other,
i.e., their separation approaches $D$.  Their contribution to the Jastrow
exponent is $J_{12}=\log(r_{12}-D)$.  Let us write the trial wave function as
$\Psi=\exp(J_{12})S$, where $S$ is well-behaved as the particles approach.
Suppose that all coordinates in the system are frozen, apart from the
separation of particles 1 and 2.  The contribution to the local energy arising
from this coordinate is
\begin{eqnarray} E_{L12} & = & -\frac{\nabla_{12}^2 \Psi}{\Psi} \nonumber \\ &
  = & |\nabla_{12} J_{12}|^2 + \nabla_{12}^2 J_{12} + 2 \nabla_{12} J_{12}
  \cdot \frac{\nabla_{12} S}{S} + \frac{\nabla_{12}^2 S}{S} \nonumber \\ & = &
  \frac{1}{r_{12}(r_{12}-D)} \left(1+2 {\bf r}_{12} \cdot \frac{\nabla_{12}
  S}{S} \right) +\frac{\nabla_{12}^2 S}{S}.  \end{eqnarray} The local energy
  diverges when hard-core particles approach one another, and the sign of the
  divergence depends on the positions of all the particles.  Hence one cannot
  remove the divergence using a two-body Jastrow factor.  By contrast, when
  two charged point particles approach one another, the divergence in the
  local energy can be removed by imposing the Kato cusp conditions via the
  two-body Jastrow factor.\cite{kato_pack}

The divergence is in principle no worse than that which occurs at any other
node.  However, there are many extra nodes introduced by the hard-core
potentials, on which the local kinetic energy diverges.  These nodes can cause
DMC population explosions,\cite{umrigar_1993} especially if three-body terms
are omitted from the Jastrow factor.

\subsection{Need for a three-body Jastrow term}

Physically, it is obvious that at high densities multi-body correlation
effects will be important: motion is only possible if the particles move
collectively.  In fact, as shown in Table \ref{table:wave_fn_comp}, three-body
Jastrow terms lower the VMC energy more than backflow.  This differs from the
behavior found in the HEG\@.\cite{ndd_2dheg_expvals} We have therefore used
three-body terms in all our calculations. If the range of the three-body terms
is restricted then the wave function becomes significantly poorer, as can be
seen in Table \ref{table:wave_fn_comp}.  Unfortunately, the need to include
long-range three-body terms in the Jastrow factor makes the calculations
expensive.

\begin{table}
\begin{center}
\begin{tabular}{lccccc} \hline \hline

Meth. &  WF & SR BF & SR 3BJ & $E$ (a.u./part.) & Var.\ (a.u.)  \\

\hline

VMC & SJ   & -- & -- & $4.8242(4)$~ & $286$ \\

VMC & SJB  & F  & -- & $4.7172(9)$~ & $242$ \\

VMC & SJB3 & T  & T  & $4.6360(2)$~ & $160$ \\

VMC & SJB3 & F  & T  & $4.6272(1)$~ & $158$ \\

VMC & SJ3  & -- & F  & $4.58672(8)$ & ~$89$ \\

VMC & SJB3 & T  & F  & $4.5201(2)$~ & ~$70$ \\

VMC & SJB3 & F  & F  & $4.5072(6)$~ & ~$65$ \\

DMC & SJ   & F  & F  & $4.4157(6)$~ & --   \\

DMC & SJB  & F  & -- & $4.4058(7)$~ & --   \\

DMC & SJB3 & F  & F  & $4.3930(8)$~ & --   \\

\hline \hline
\end{tabular}
\caption{Non-twist-averaged VMC and DMC results with different wave functions
(``WF'') for a paramagnetic fluid of $N=26$ hard-core particles of diameter
$D=1$ a.u.  The ``SR BF'' column specifies whether or not the backflow
function was restricted to be short-ranged (cutoff length 2.5 a.u.), while the
``SR 3BJ'' column specifies whether or not the three-body Jastrow terms were
restricted to be short-ranged (cutoff length 2.5 a.u.).  Where the cutoff
lengths were not restricted, they approached the radius of the circle
inscribed in the Wigner-Seitz cell of the simulation cell.  ``T'' and ``F''
denote true and false, while ``Var.''\ is the variance of the local energy.
\label{table:wave_fn_comp}}
\end{center}
\end{table}

\subsection{Spin-dependence in paramagnetic phases}

We used separate two-body Jastrow and two-body backflow terms for
parallel-spin and antiparallel-spin pairs of particles in all our
calculations.  We considered two possible spin-dependences for the three-body
terms: either (A) using the same three-body term for all triples of particles
or (B) using separate three-body terms for (i) triples involving three
particles of the same spin and (ii) triples involving two particles of one
spin and one particle of the opposite spin.  VMC results obtained using these
two different possibilities are shown in Table \ref{table:spin_dep}.  It is
clear that allowing different three-body terms for different spin
configurations lowers the variational energy significantly (although not by
nearly as much as lifting the restriction on the range of the three-body
terms) and hence we have used spin-dependence B in all our production
calculations.

\begin{table}
\begin{center}
\begin{tabular}{lccc} \hline \hline

Spin-dep. & SR BF \& 3BJ & Energy (a.u./part.) & Var.  (a.u.) \\

\hline

A & T & $4.640(1)$~ & $165$ \\

B & T & $4.6360(2)$ & $160$ \\

A & F & $4.534(1)$~ & ~$76$ \\

B & F & $4.5072(6)$ & ~$65$ \\

\hline \hline
\end{tabular}
\caption{Non-twist-averaged VMC results with different spin dependences for a
paramagnetic fluid of $N=26$ hard-core particles of diameter $D=1$ a.u.  The
``SR BF \& 3BJ'' column specifies whether or not the backflow function and
three-body Jastrow terms were restricted to be short-ranged (cutoff length 2.5
a.u.). Where the cutoff lengths were not restricted, they approached the
radius of the circle inscribed in the Wigner-Seitz cell of the simulation
cell. ``T'' and ``F'' denote true and false, while ``Var.''\ is the variance
of the local energy.\label{table:spin_dep}}
\end{center}
\end{table}

\subsection{Relative accuracy of wave functions for different phases}

The variance of the energy is zero if the trial wave function is an
eigenfunction of the Hamiltonian.  As shown in Table \ref{table:wf_accuracy},
the variance per particle is significantly lower for the ferromagnetic fluid
than the paramagnetic fluid, indicating that the trial wave function is more
accurate for the former than the latter.  The variance per particle is similar
for ferromagnetic fluids and crystals near the transition density.  Hence, if
anything, our results are biased in favor of ferromagnetic phases.

\begin{table}
\begin{center}
\begin{tabular}{lcc} \hline \hline

$D$ (a.u.) & Phase & Var.\ per part.\ (a.u.) \\

\hline

$1$    & Para.\ fluid    & $2.49$~ \\

$1$    & Ferro.\ fluid   & $1.76$~ \\

$1$    & Ferro.\ crystal & $0.736$ \\

$0.88$ & Para.\ fluid    & $1.34$~ \\

$0.88$ & Ferro.\ fluid   & $0.616$ \\

$0.88$ & Ferro.\ crystal & $0.494$ \\

$0.63$ & Para.\ fluid    & $0.385$ \\

$0.63$ & Ferro.\ fluid   & $0.147$ \\

$0.63$ & Ferro.\ crystal & $0.228$ \\

\hline \hline
\end{tabular}
\caption{Non-twist-averaged VMC results for paramagnetic fluids with $N=26$
hard-core particles and ferromagnetic fluids with $N=25$ hard-core
particles. \label{table:wf_accuracy}}
\end{center}
\end{table}

\subsection{Nature of phase transition}

We have looked for a first-order phase transition by comparing fixed-node DMC
energies with fluid and crystal orbitals, relying on the fixed-node
approximation to impose the symmetry of the phase on the wave function.
However, it is possible that there could actually be a continuous transition
from fluid to crystal behavior.  The fact that the ``fluid'' wave function
tries to become crystal-like at high density and the ``crystal'' wave function
tries to become fluid-like at low density (see Sec.\ \ref{sec:expvals})
supports this view.  Nevertheless, even if this is the case, our calculations
determine the region in which crystallization is expected to take place, and
demonstrate that a ferromagnetic fluid phase is unlikely to occur.

\section{Finite-size effects \label{sec:finite_size}}

According to the simple theory given in Sec.\ \ref{sec:parallel_bcs}, the
long-range parallel-spin two-body Jastrow exponent is approximately given by
\begin{equation} u_{\alpha \alpha}(r) = \log(1-D^2/r^2) \approx
-D^2/r^2 + O(r^{-4}).
\end{equation} The 2D Fourier transform of
the leading term does not have a power-law behavior.  In fact numerical tests
show that $\lim_{\alpha \rightarrow 0} \int r^{-2} \exp(i{\bf k}\cdot {\bf r}
- \alpha r) \, d{\bf r}$ diverges logarithmically at small $k$.  So, for small
$k$, $u_{\alpha \alpha}(k)$ may be written as $-cD^2 \log(k)$, where $c$ is a
positive constant.

The Chiesa-Holzmann-Martin-Ceperley\cite{fin_chiesa} approximation to the
long-range two-body finite-size correction to the kinetic energy in 2D is
\begin{eqnarray} \Delta T & = & - \sum_\alpha
\frac{N_\alpha}{4(2\pi)^2} \int_0^Q 2\pi k \times k^2 u_{\alpha \alpha}(k) \,
dk \nonumber \\ & = & -\frac{N}{8 \pi} \int_0^Q k^3 u_{\alpha \alpha}(k) \, dk,
\end{eqnarray} where $Q=2\sqrt{\pi/A}$ is the radius of the circle in ${\bf
k}$-space with area $(2\pi)^2/A$ and $A=\pi N$ is the area of the simulation
cell.  Inserting $u_{\alpha \alpha}(k)=-cD^2 \log(k)$ gives
\begin{eqnarray} \Delta T & \approx & \frac{\pi N c D^2}{2A^2} \left[
\log \left(2\sqrt{\pi/A} \right)-4\pi^2/A^2 \right] \nonumber \\ & = & O[D^2
N^{-1}\log(N)]. \end{eqnarray} Hence, ignoring the logarithmic factor, the
leading-order correction to the energy per particle due to the neglect of
long-range two-body correlations falls off as $O(N^{-2})$.  This is much more
rapid than the analogous result for the 2D HEG\@.\cite{ndd_fs} The
leading-order correction is positive.  Unsurprisingly, however, two-body
finite-size errors get rapidly more severe as $D$ increases.

\begin{figure}
\begin{center}
\includegraphics[clip,scale=0.9]{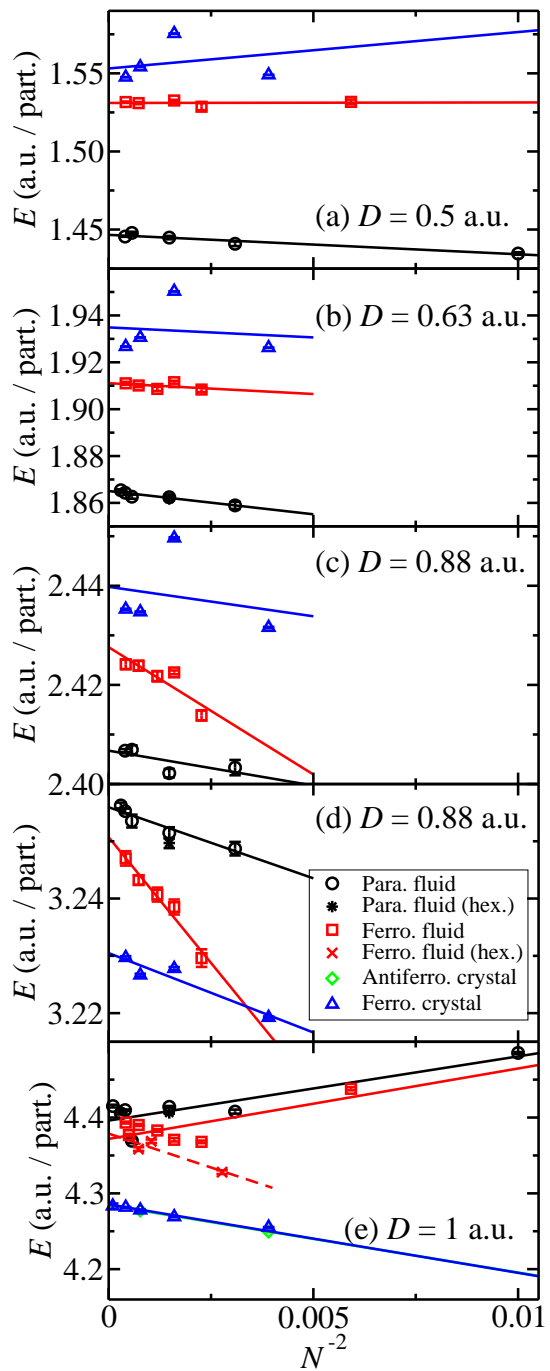}
\caption{(Color online) DMC energy $E$ against the reciprocal of the square of
the system size $N^{-2}$ for (a) $D=0.5$ a.u., (b) $D=0.63$ a.u., (c) $D=0.75$
a.u., (d) $D=0.88$ a.u., and (e) $D=1$ a.u.  \label{fig:E_v_N}}
\end{center}
\end{figure}

We have carried out numerical tests which show that the bias due to twist
averaging in the canonical ensemble in two dimensions falls off as
$O(N^{-3/2})$, but with a small prefactor.  Residual single-particle errors in
the canonical-ensemble twist-averaged fluid energies are small compared with
the effect of twist averaging, and the difference between the twist-averaged
and non-twist-averaged energies is only about 0.02 a.u.\ per particle at $D=1$
a.u.\ for $N=10$ and $26$.  Hence the $O(N^{-2})$ behavior due to the neglect
of long-range two-body correlations dominates the systematic error for all
the hard-core diameters $D$ that we have studied, as can be seen in Fig.\
\ref{fig:E_v_N}.

\begin{figure}
\begin{center}
\includegraphics[clip,scale=0.3]{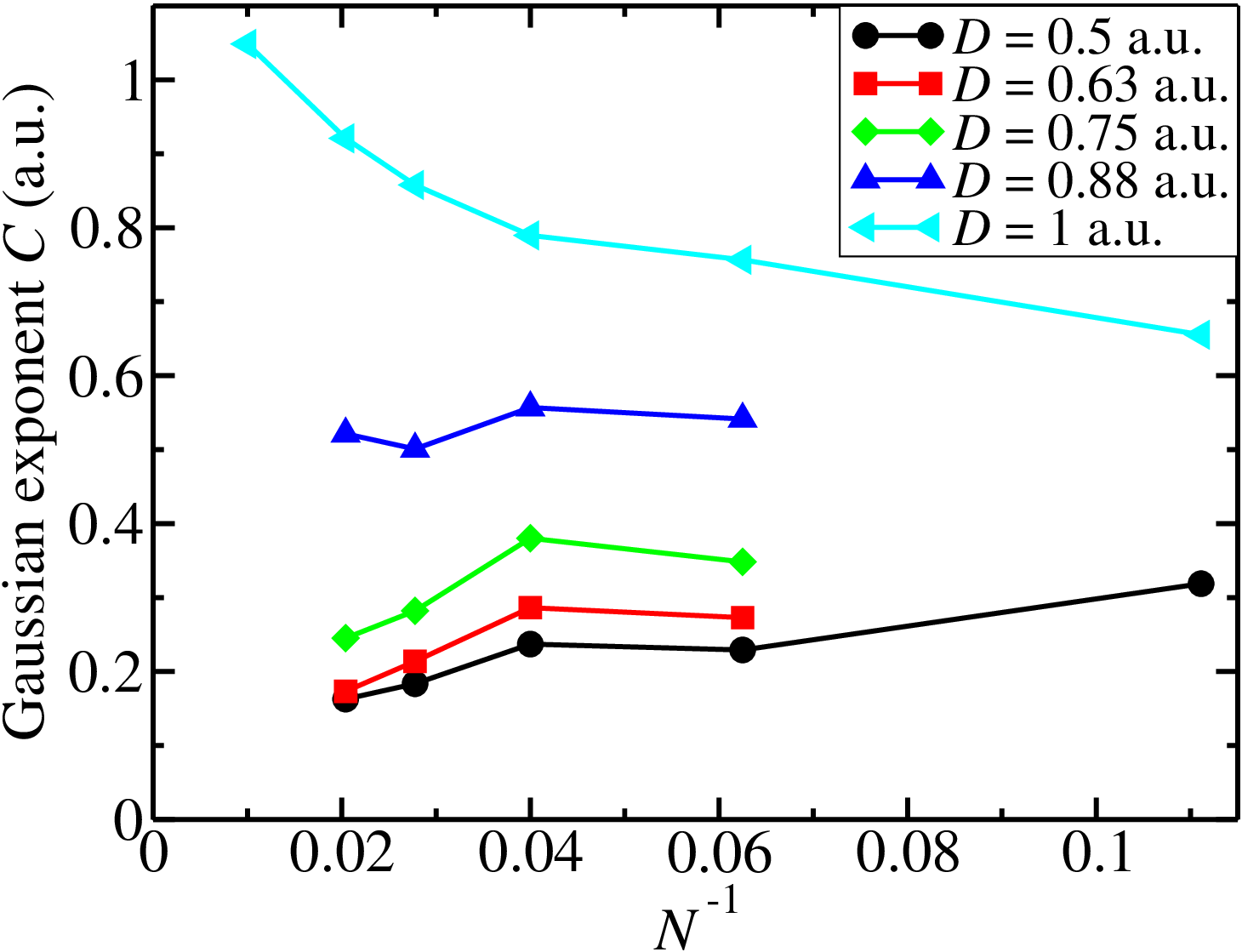}
\caption{(Color online) Optimal Gaussian exponent $C$ for the ferromagnetic
crystal phase against the reciprocal of system size
$N^{-1}$. \label{fig:gauss_exponents}}
\end{center}
\end{figure}

Several data points in Fig.\ \ref{fig:E_v_N} are outliers.  This nonsystematic
behavior appears to be a genuine finite-size effect.  The finite-size
``noise'' in the crystal energies is more severe at low densities, while the
finite-size noise in the fluid energies is more severe at very high densities.
The theory of two-body finite-size effects developed above breaks down for
fluids at high density and crystals at low density: the sign of the bias is
wrong.  These cases are pathological in various respects.  At very high
densities, the fluid energies obtained with the same number of particles in
different-shaped cells disagree, although they extrapolate to the same value
in the limit of infinite system size.  In fact at very high density it is
geometrically impossible to fit finite fluids in some cell shapes.  As can be
seen in Fig.\ \ref{fig:gauss_exponents}, the Gaussian exponent $C$ of the
crystal is not well-behaved for $D\leq 0.75$, suggesting that the crystal is
becoming unstable.  As shown in Fig.\ \ref{fig:crys_mom_dist}, the momentum
density is developing an edge, leading to substantial single-particle
finite-size errors.

As shown in Table \ref{table:wave_fn_comp}, restricting the range of the
three-body term raises the variational energy significantly.  Three-body
correlations are therfore long-ranged.  So there must be three-body
finite-size errors in the fluid and crystal energies obtained in finite
simulation cells. We assume they are a ``random'' error about the systematic
finite-size bias due to two-body finite-size effects.  We have therefore
obtained several $E(N)$ data points for each phase at each density in order to
average out the ``noise'' when extrapolating to infinite system size.

\section{Phase diagram \label{sec:phase_diagram}}

\begin{figure}
\begin{center}
\includegraphics[clip,scale=0.3]{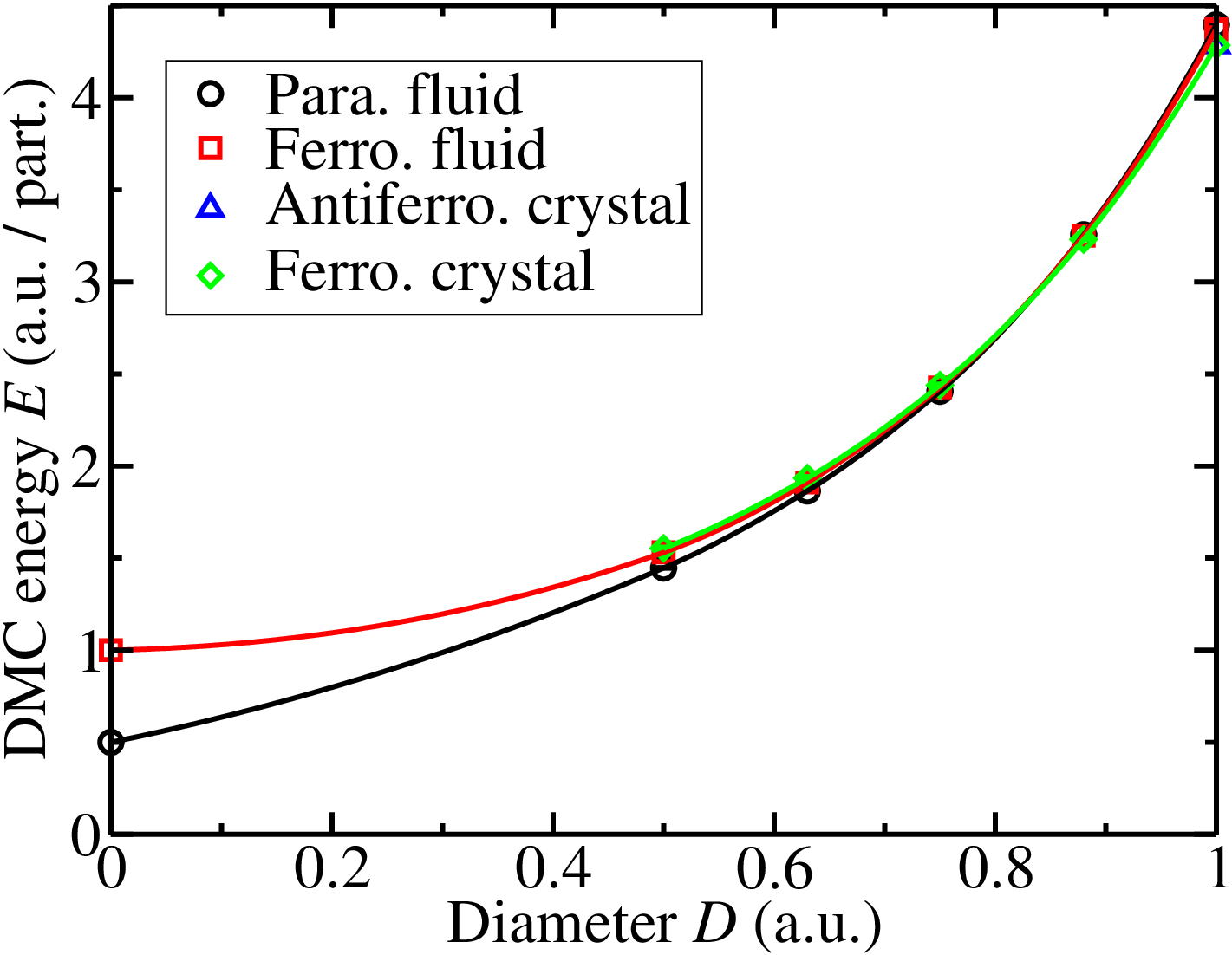}
\caption{(Color online) DMC energy $E$ against hard-core diameter $D$.  Note
that, when $D=0$ the energies of the paramagnetic and ferromagnetic fluids are
0.5 and 1 a.u.\ per particle, respectively. \label{fig:E_v_D}}
\end{center}
\end{figure}

\begin{figure}
\begin{center}
\includegraphics[clip,scale=0.3]{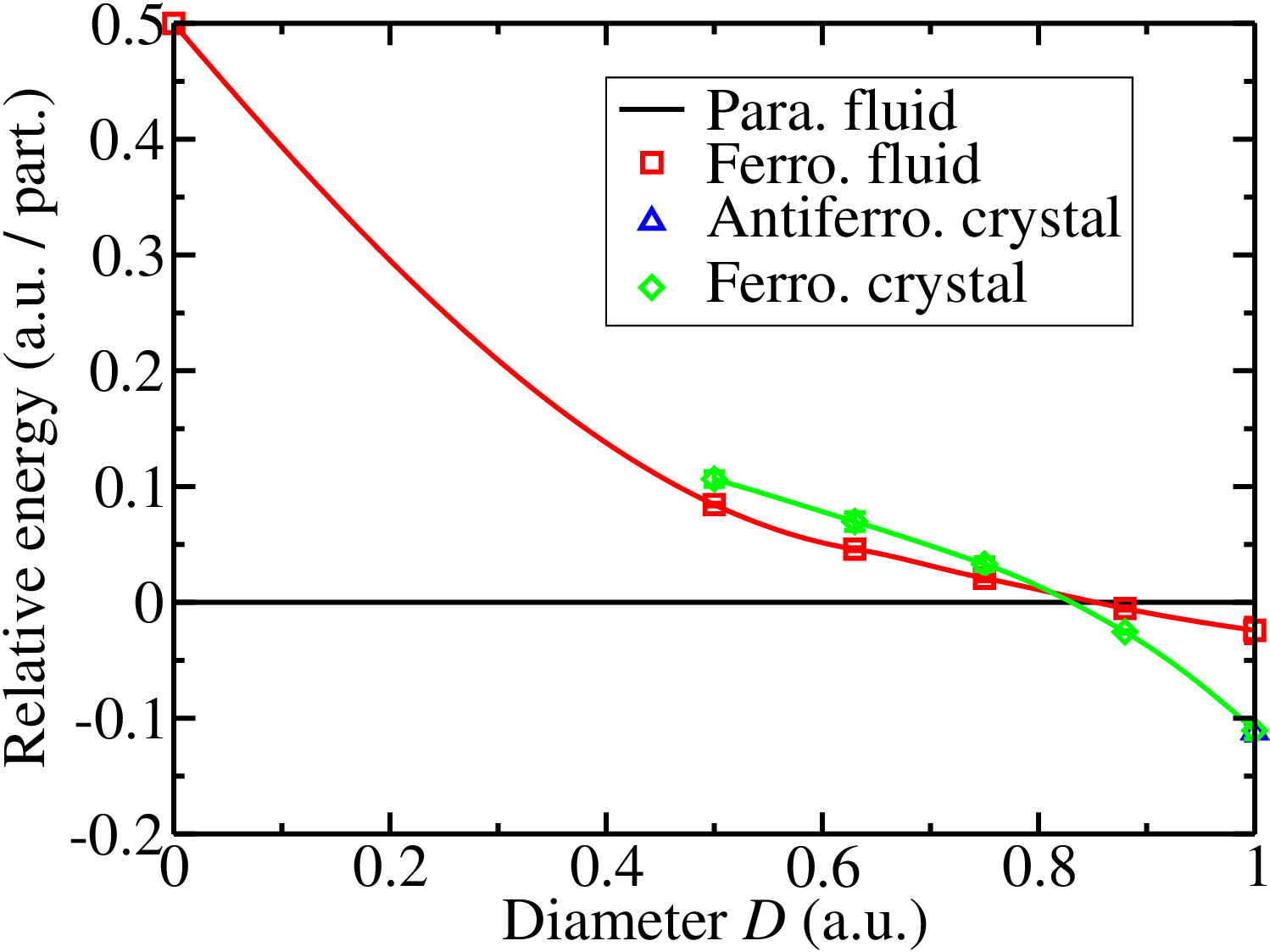}
\caption{(Color online) DMC energy of the different phases relative to the
energy of the paramagnetic fluid. \label{fig:relE_v_D}}
\end{center}
\end{figure}

The energies of the different phases are compared in Fig.\ \ref{fig:E_v_D} and
are plotted relative to the energy of the paramagnetic fluid phase in Fig.\
\ref{fig:relE_v_D}.  The continuous curves shown are Akima spline
interpolations between the DMC energies.  (Akima spline interpolation is
stable to the presence of outliers in the data.\cite{akima}) It can be seen
that crystallization takes place when $D=0.83$ a.u., leaving no region of
stability for a ferromagnetic fluid.  (Recall that our calculations are, if
anything, biased in favor of the ferromagnetic fluid.)  Thus our calculations
rule out the possibility of an itinerant ferromagnetic fluid phase in a 2D gas
of particles with only hard-core interactions.  For those values of $D$ for
which the crystal has the lowest energy, the energy difference between the
antiferromagnetic and ferromagnetic states is insignificant, showing that
exchange interactions in the crystalline phase are negligible.

\section{Other properties of hard-core gases \label{sec:expvals}}

\subsection{Pair-correlation function\label{sec:pcf}}

\begin{figure}
\begin{center}
\includegraphics[clip,scale=0.27]{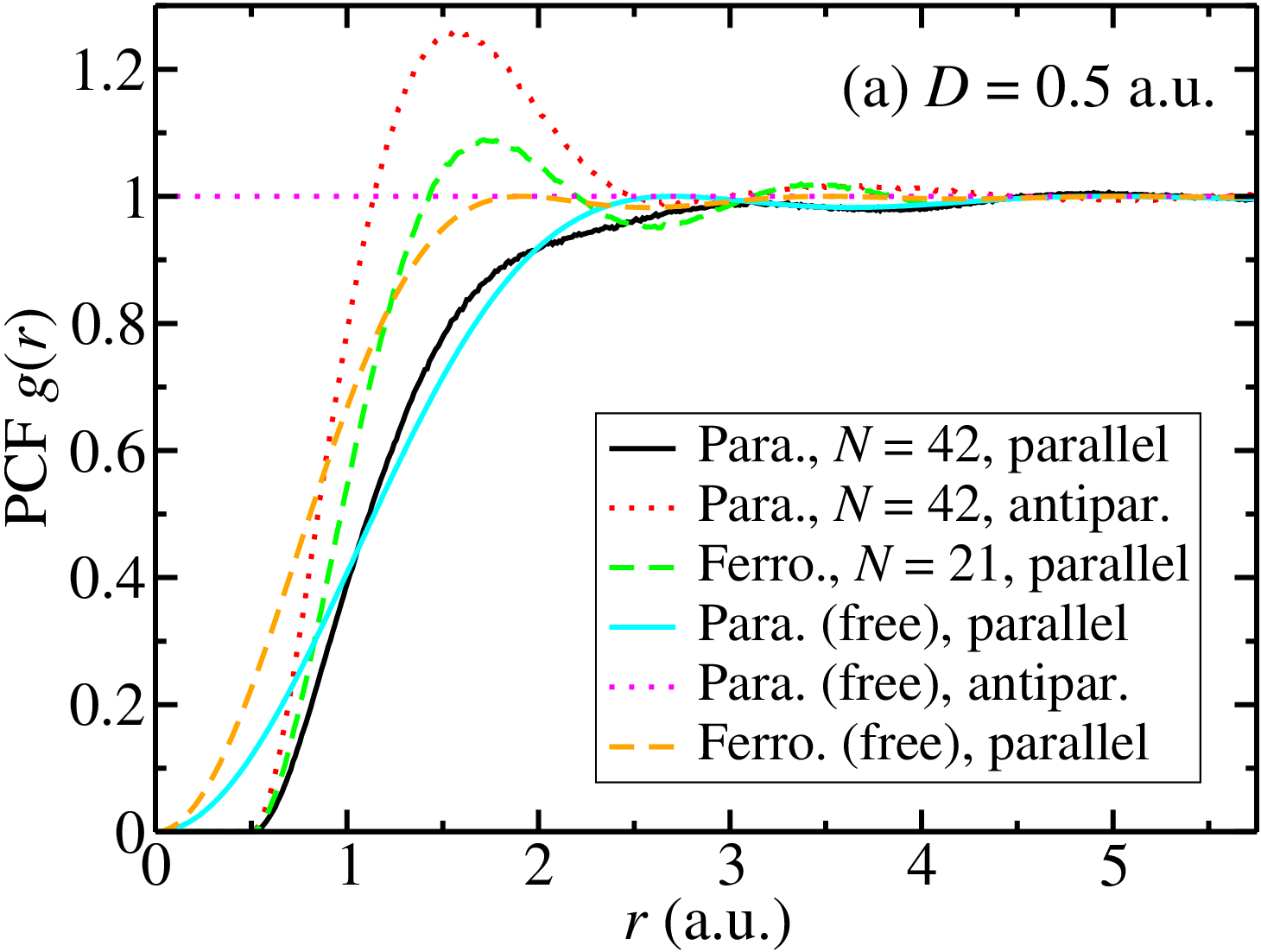} \\
\includegraphics[clip,scale=0.27]{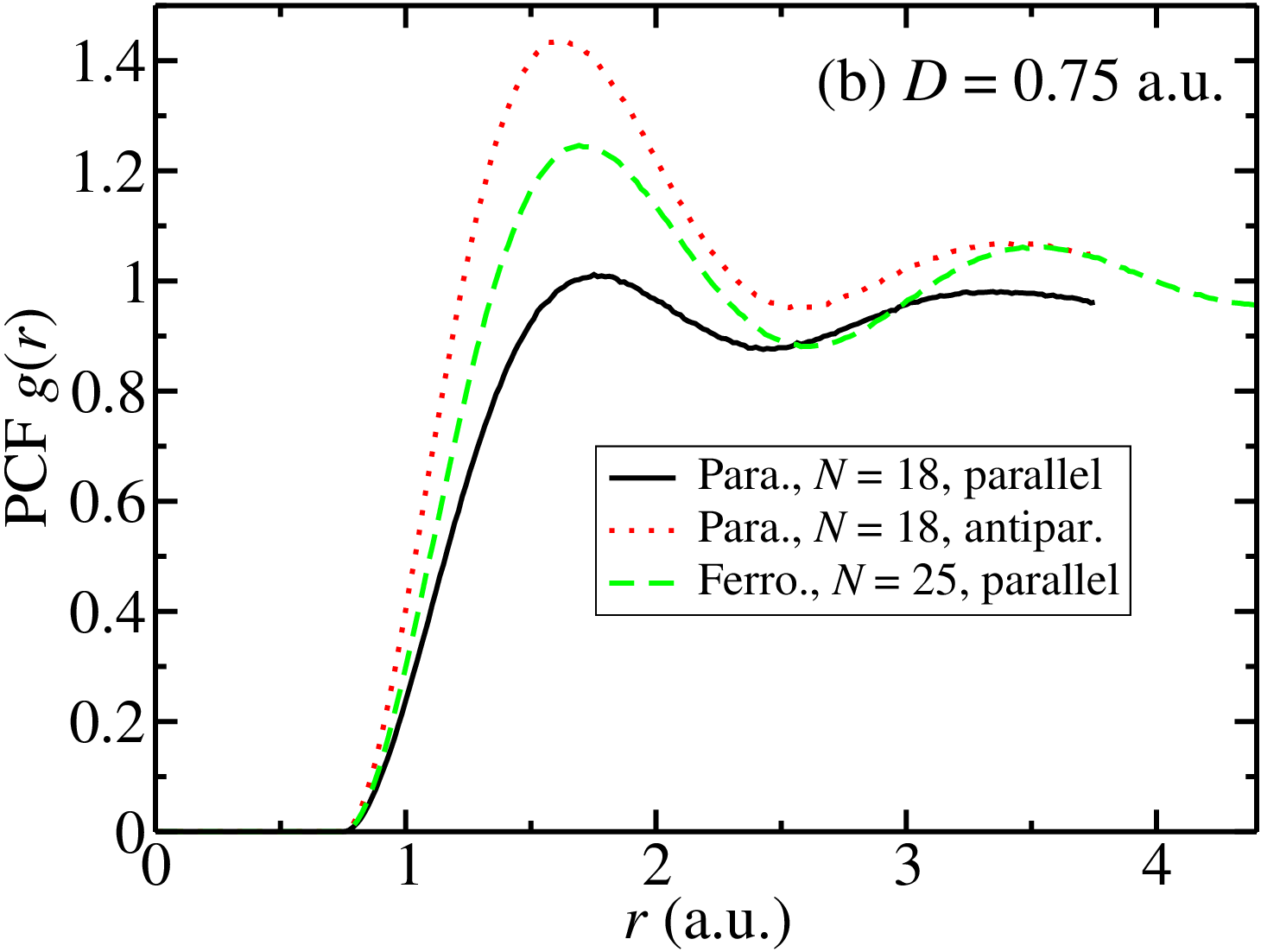} \\
\includegraphics[clip,scale=0.27]{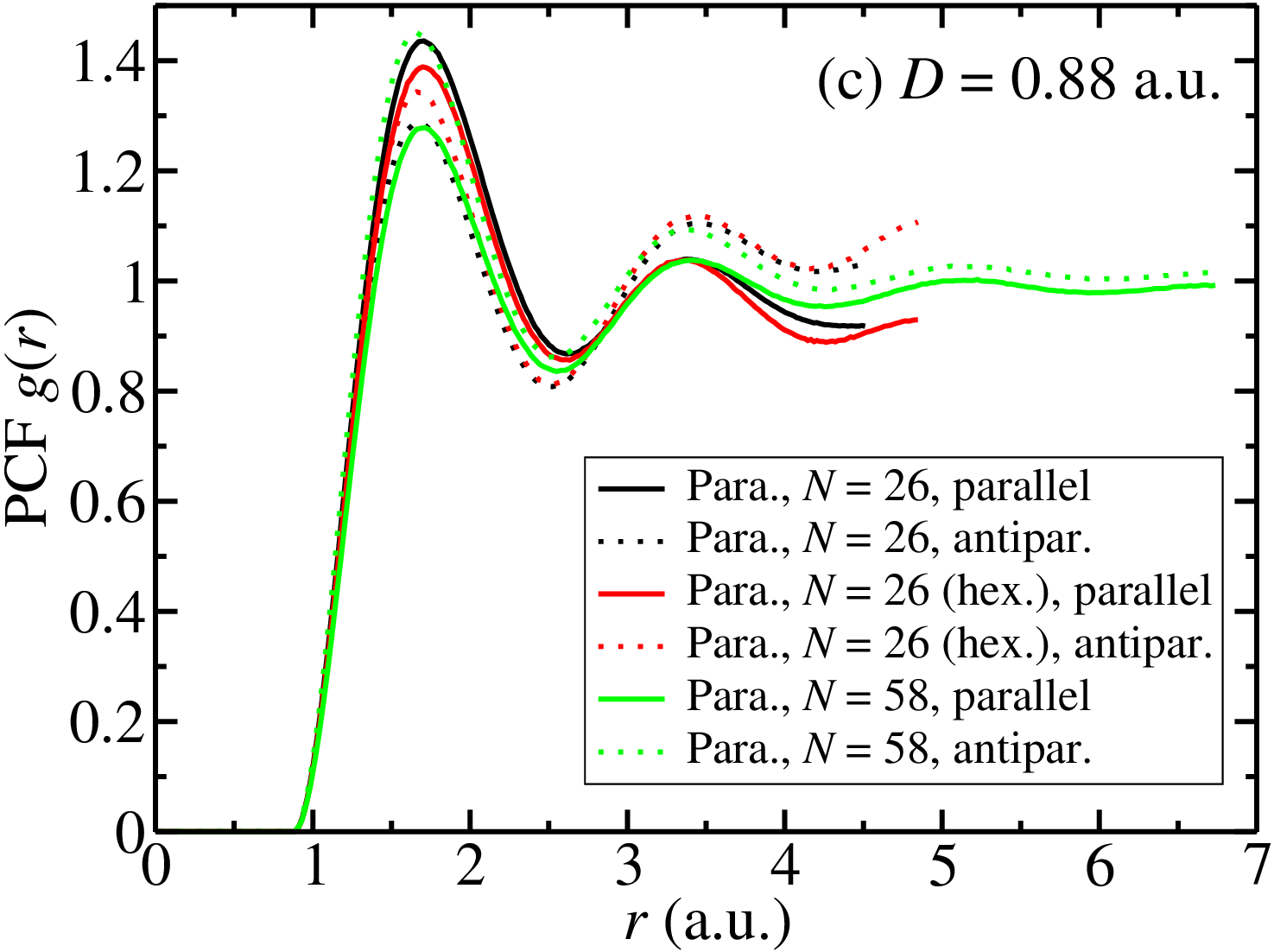} \\
\includegraphics[clip,scale=0.27]{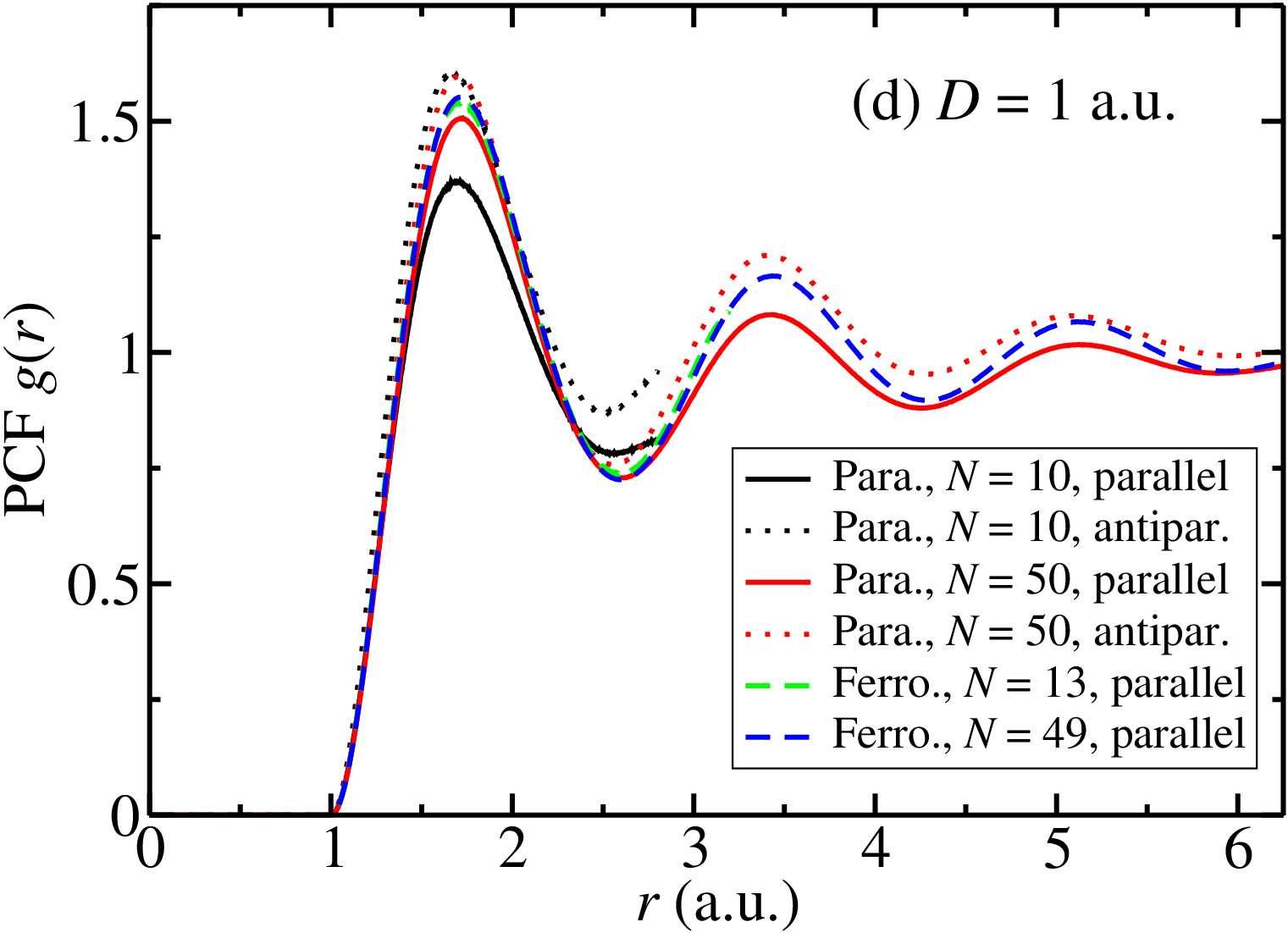}
\caption{(Color online) VMC PCFs $g(r)$ of hard-core systems at (a) $D=0.5$
  a.u., (b) $D=0.75$ a.u., (c) $D=0.88$ a.u., and (d) $D=1$ a.u., for
  different system sizes $N$. \label{fig:hard_pcf}}
\end{center}
\end{figure}

We have calculated the PCFs of the fluid phases of the hard-core gas, and our
results are shown in Fig.\ \ref{fig:hard_pcf}.  The PCFs, which were obtained
without twist averaging, are not well-converged with respect to system size at
high density: the finite-size errors are oscillatory.  Nevertheless, we can
make some qualitative comments about the physics revealed by the PCF\@.

The distance between the peaks of the PCF obtained in square cells at $D=1$
a.u.\ is $\sqrt{\pi}$, which arises from the square-cell geometry.  The
difference between the PCFs in hexagonal and square cells is still significant
at $D=0.88$ a.u.\ and $N=26$.  The antiparallel-spin PCF evolves slowly
towards $g(r)=1$ (the free-particle result) as $D$ is reduced.  The size of
the exchange-correlation hole and the ripples in the parallel-spin PCF are
determined by the Fermi wave vector $k_F$ when $D$ is small.

\subsection{Static structure factor}

\begin{figure}
\begin{center}
\includegraphics[clip,scale=0.3]{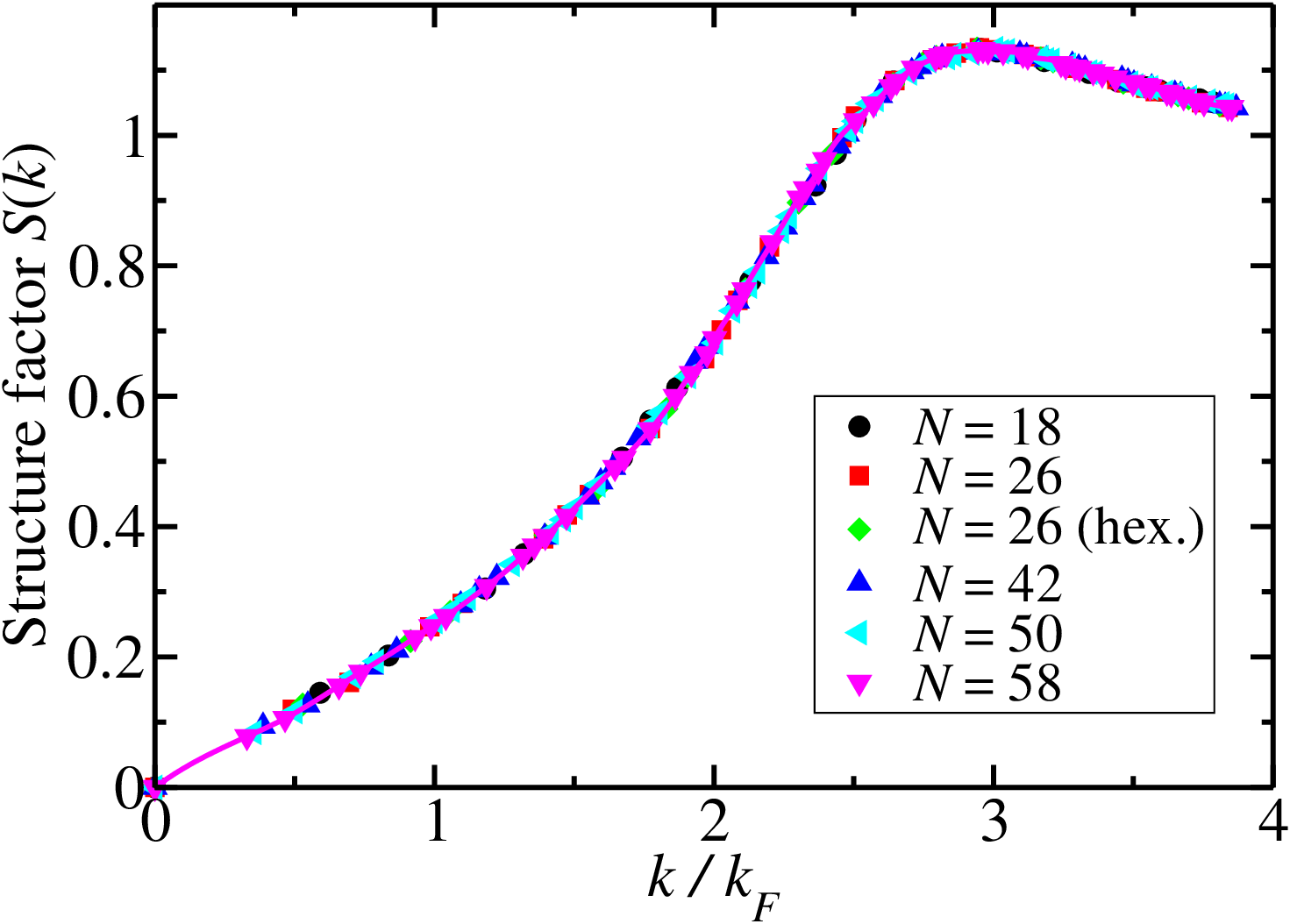}
\caption{(Color online) VMC structure factor of the paramagnetic fluid at
  $D=0.63$ a.u.\ for a range of system sizes $N$. \label{fig:para_sf_D0.63}}
\end{center}
\end{figure}

The static structure factor of a paramagnetic fluid at $D=0.63$ a.u.\ for
various values of $N$ is plotted in Fig.\ \ref{fig:para_sf_D0.63}.  The
structure factor is well converged with respect to $N$.

\begin{figure}
\begin{center}
\includegraphics[clip,scale=0.3]{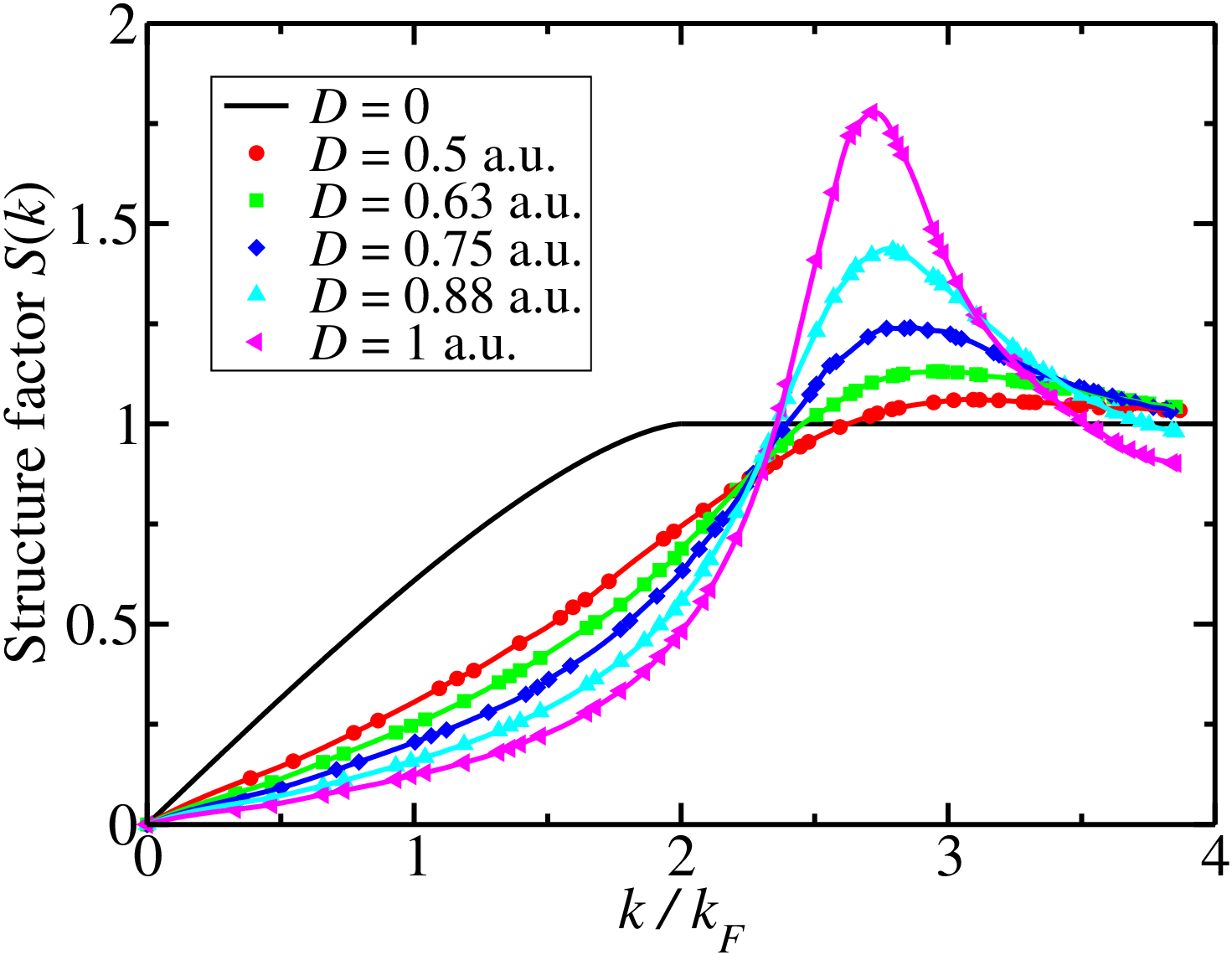}
\caption{(Color online) VMC structure factors of paramagnetic fluids.  For
$D=0$ the analytic result for a paramagnetic free-particle gas is
given. \label{fig:para_SF}}
\end{center}
\end{figure}

\begin{figure}
\begin{center}
\includegraphics[clip,scale=0.3]{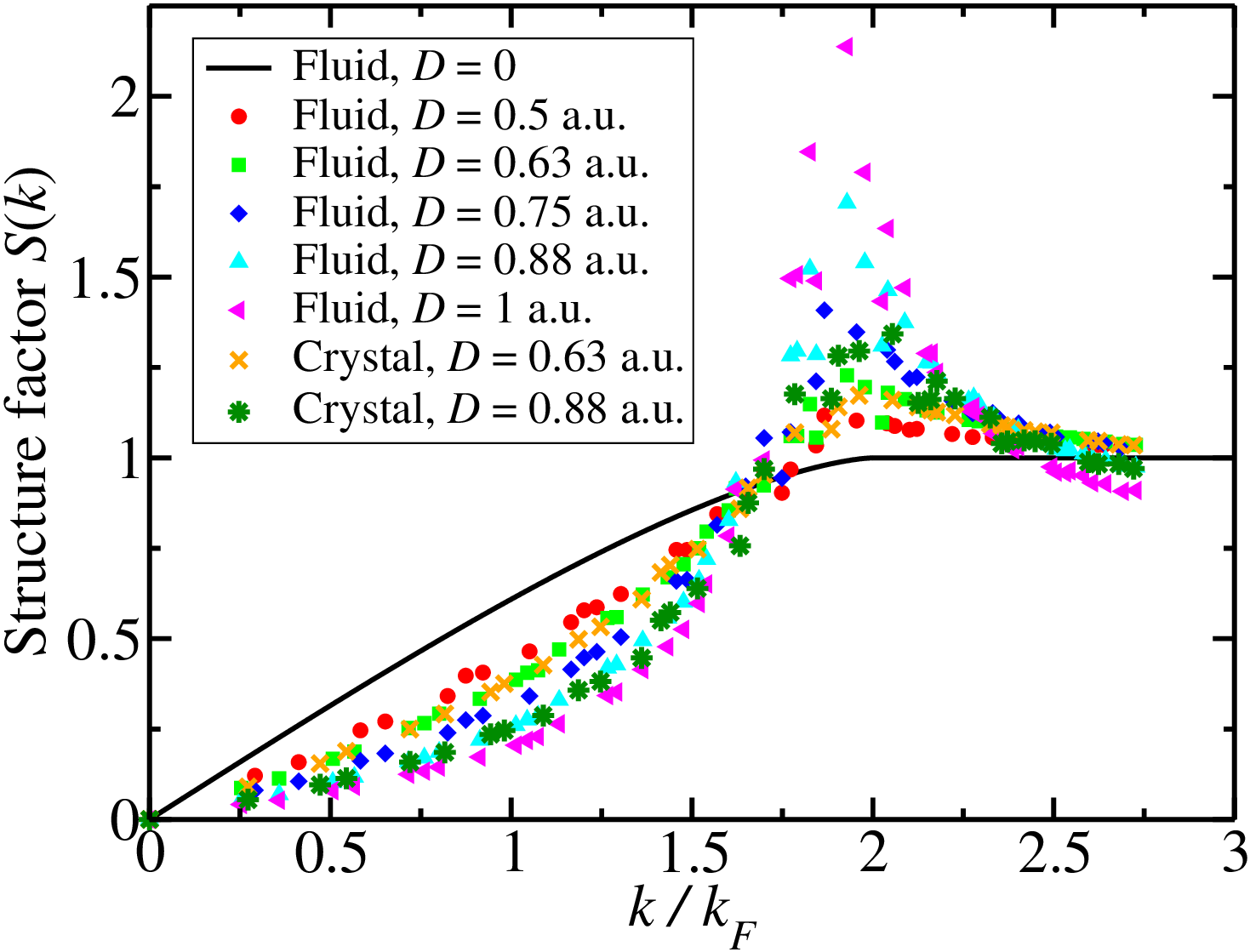}
\caption{(Color online) VMC structure factors of ferromagnetic fluids and
crystals.  The $D=0$ result is the analytic structure factor for a
ferromagnetic free-particle gas. \label{fig:ferro_SF}}
\end{center}
\end{figure}

VMC-calculated structure factors for paramagnetic and ferromagnetic hard-core
gases are shown in Figs.\ \ref{fig:para_SF} and \ref{fig:ferro_SF},
respectively.  Finite-size ``noise'' is much worse for ferromagnetic phases.
The fluid and crystal structure factors are similar, especially at low
densities.

\subsection{Momentum density}

\subsubsection{Results for fluids}

\begin{figure}
\begin{center}
\includegraphics[clip,scale=0.3]{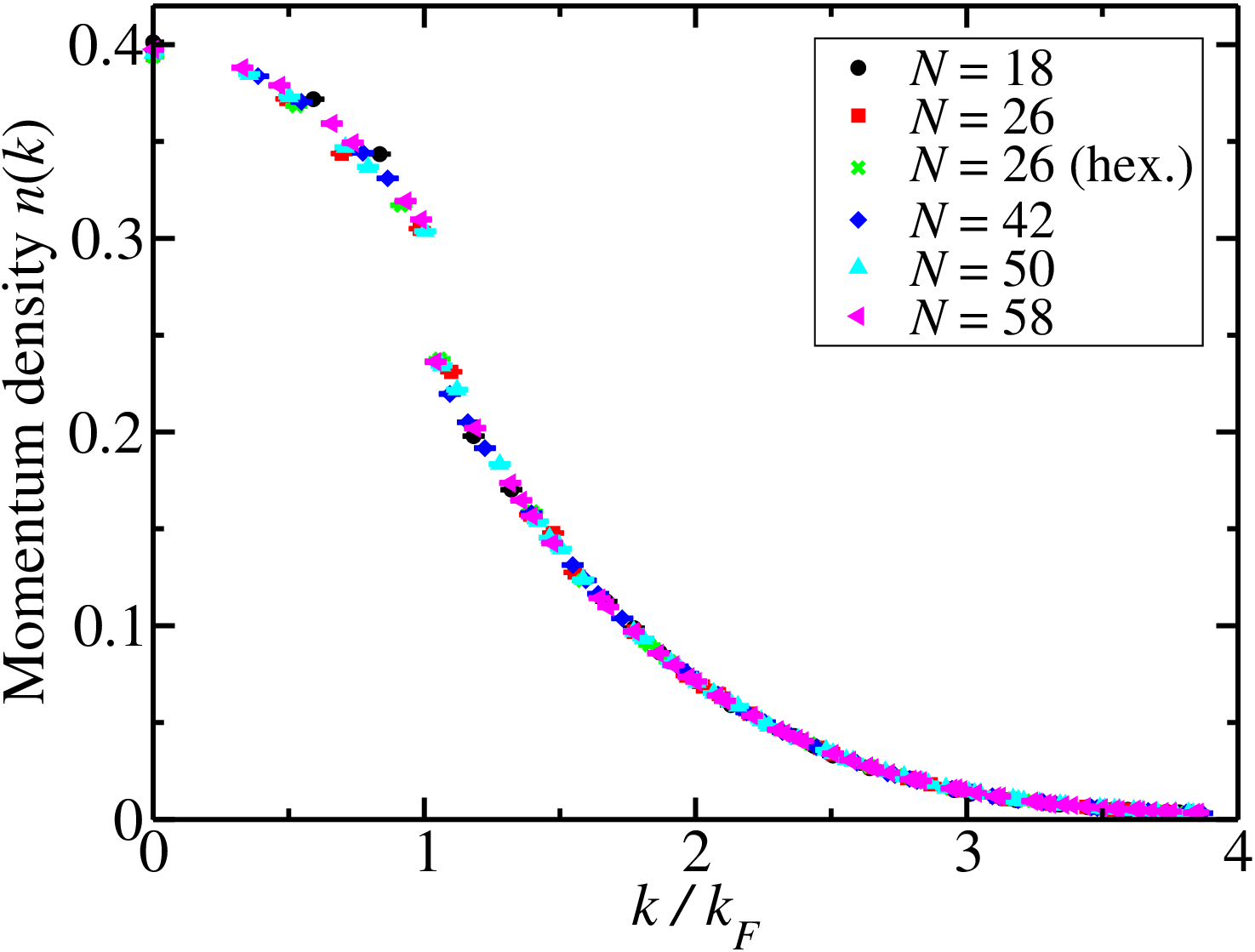}
\caption{(Color online) VMC momentum density of the paramagnetic fluid at
  $D=0.88$ a.u.\ at different system sizes $N$. \label{fig:mom_den_D0.88}}
\end{center}
\end{figure}

\begin{figure}
\begin{center}
\includegraphics[clip,scale=0.3]{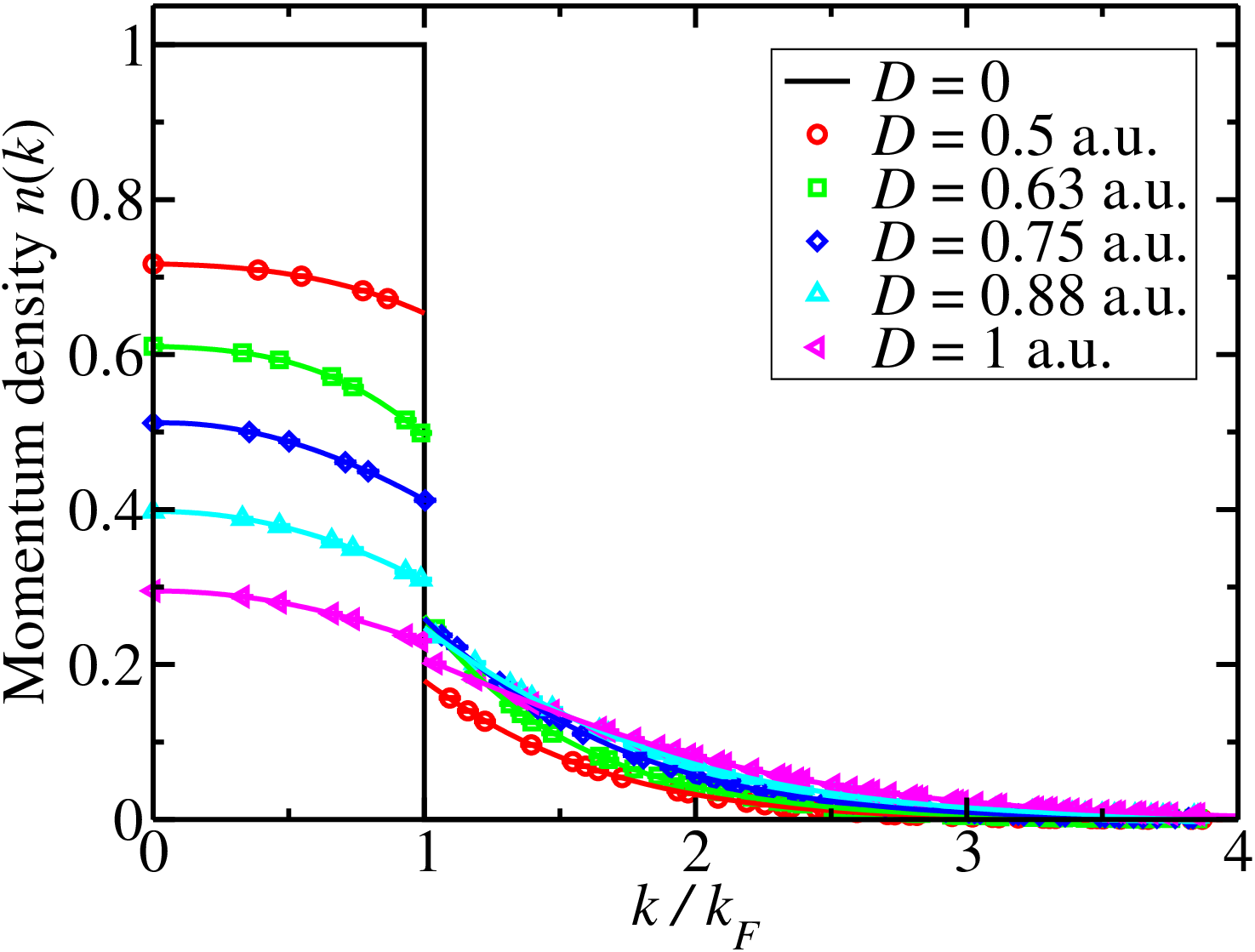}
\caption{(Color online) VMC momentum densities of paramagnetic fluids at
different hard-core diameters $D$. \label{fig:mom_den_para}}
\end{center}
\end{figure}

\begin{figure}
\begin{center}
\includegraphics[clip,scale=0.3]{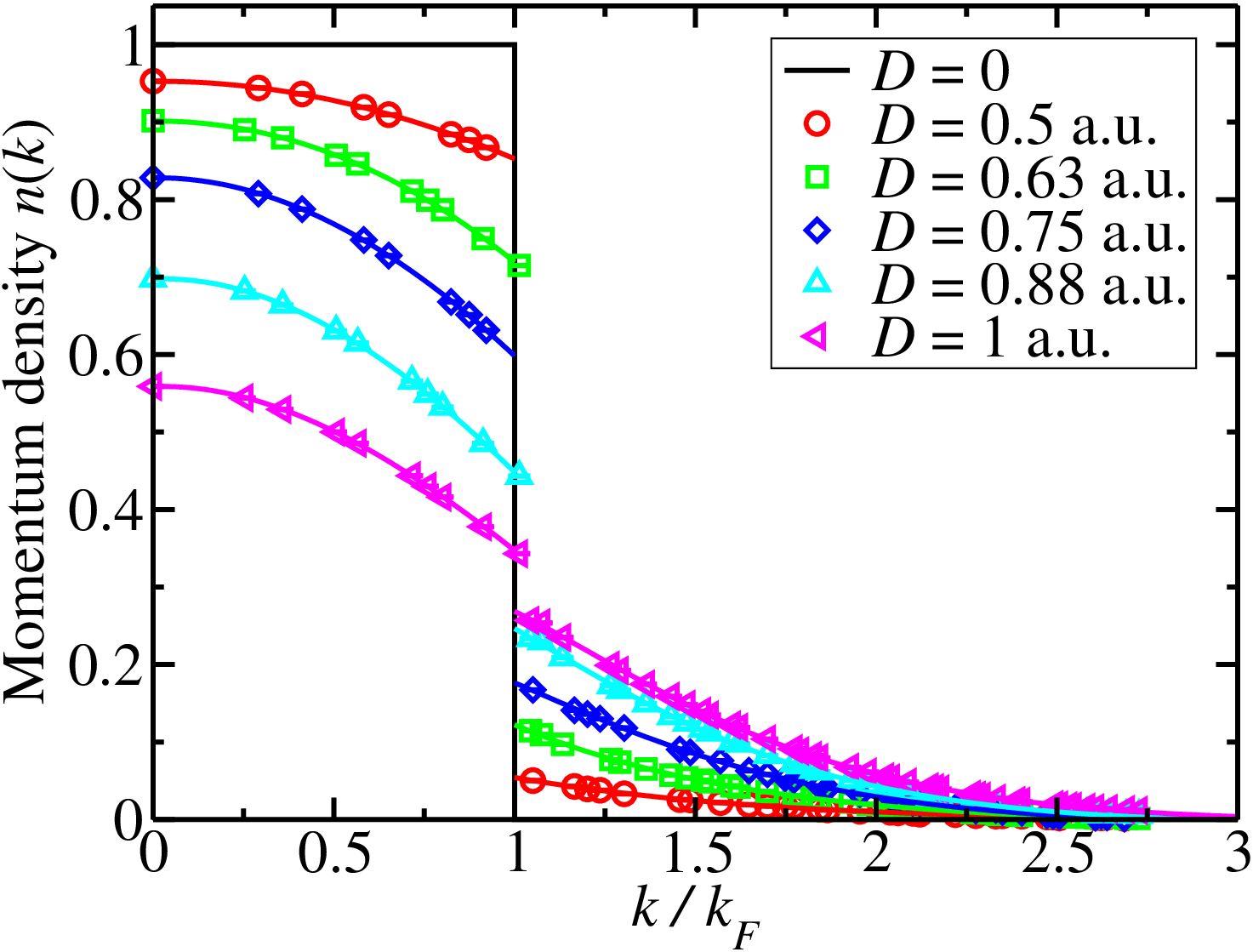}
\caption{(Color online) VMC momentum densities of ferromagnetic fluids at
different hard-core diameters $D$. \label{fig:mom_den_ferro}}
\end{center}
\end{figure}

The momentum density shown in Fig.\ \ref{fig:mom_den_D0.88} is reasonably well
converged with respect to $N$.  VMC results for the momentum densities of
paramagnetic and ferromagnetic hard-core systems are shown in Figs.\
\ref{fig:mom_den_para} and \ref{fig:mom_den_ferro}, respectively.

\subsubsection{Renormalization factors}

\begin{figure}
\begin{center}
\includegraphics[clip,scale=0.3]{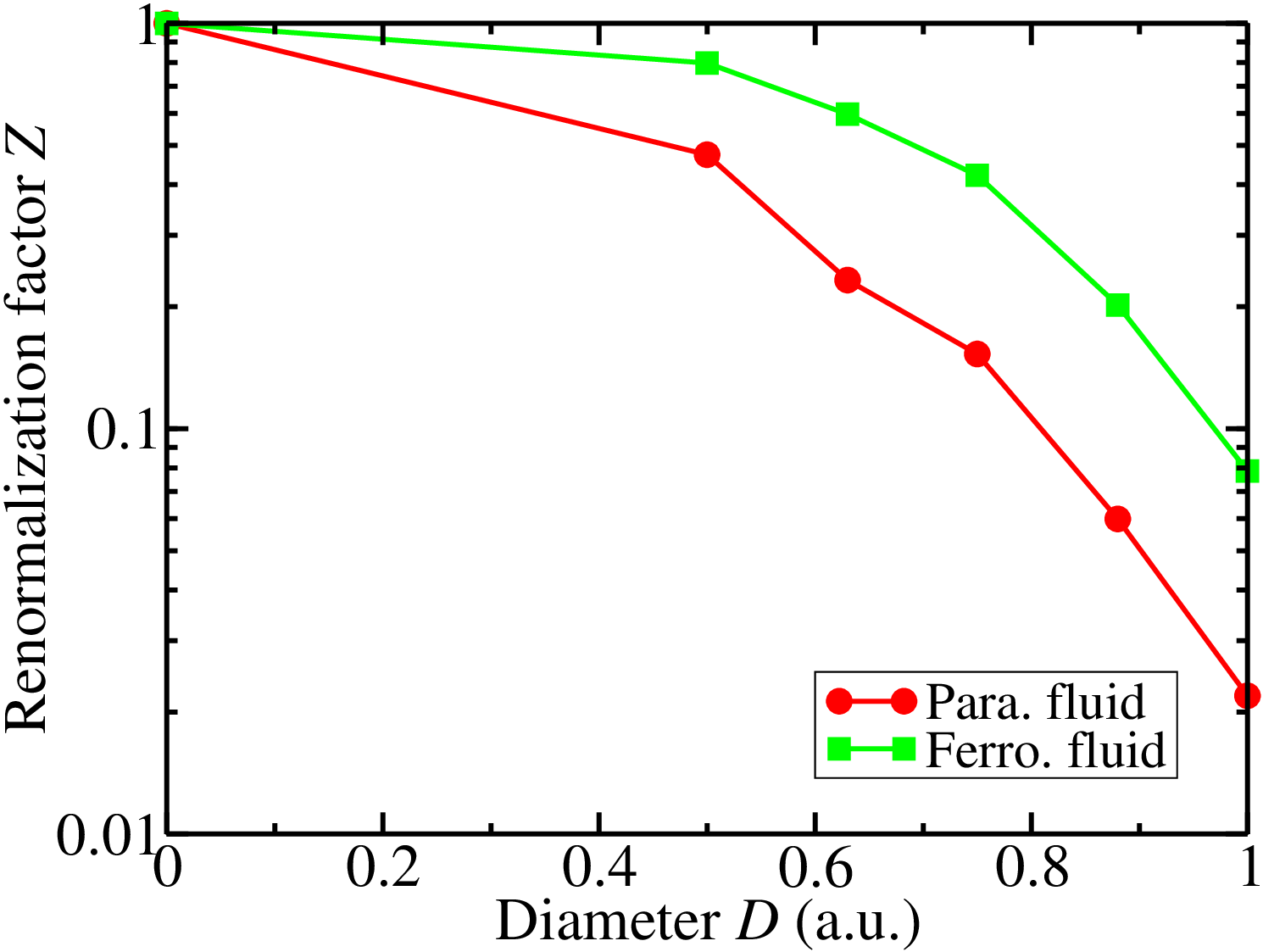}
\caption{(Color online) Renormalization factor $Z$ against $D$, from fits to
  the momentum density. \label{fig:Z_v_D}}
\end{center}
\end{figure}

The renormalization factor---the discontinuity $Z$ in the momentum density at
the Fermi edge---is plotted against hard-core diameter in Fig.\
\ref{fig:Z_v_D}.  The finite-size errors in $Z$ are oscillatory, so we have
averaged over the results obtained for different system sizes $N$.

\subsubsection{Results for crystals}

\begin{figure}
\begin{center}
\includegraphics[clip,scale=0.3]{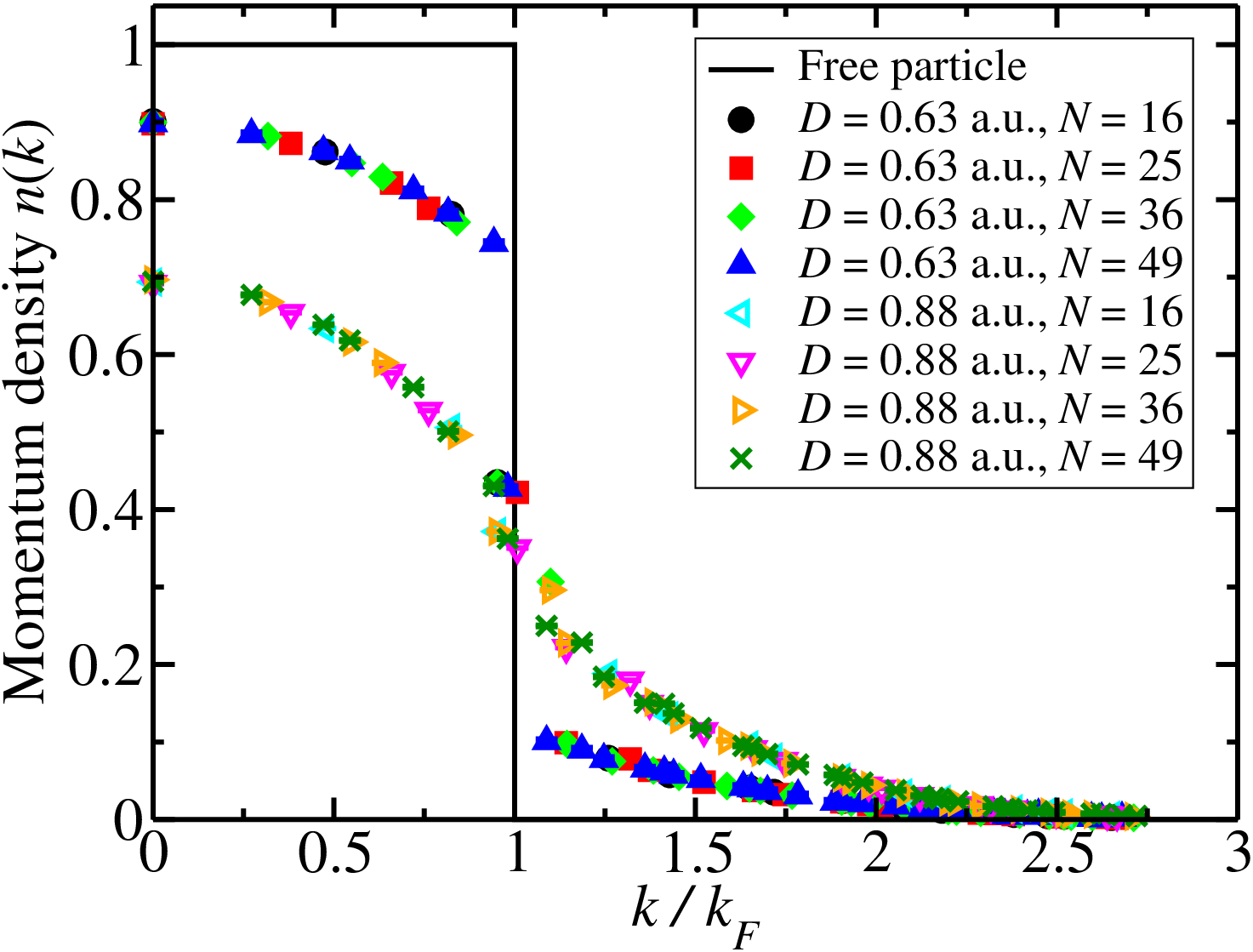}
\caption{(Color online) VMC momentum densities of ferromagnetic
crystals. \label{fig:crys_mom_dist}}
\end{center}
\end{figure}

The momentum densities of ferromagnetic hard-core crystals are shown in Fig.\
\ref{fig:crys_mom_dist}. This figure strongly confirms the conclusion that the
crystal is unstable at $D=0.63$ a.u.:\ the ``crystal'' momentum density
develops a near-discontinuity at the Fermi wave vector.

\section{Weak, long-range interactions \label{sec:attractive_interactions}}

\subsection{Form of the interaction}

Our results show that, close to the transition from fluid to crystal, the
energies of the paramagnetic and ferromagnetic fluids are very finely
balanced, albeit with the paramagnetic fluid always lying lower in energy. It
is interesting to ask whether a small change in the two-body potential might
alter the relative stabilities of these two fluid phases and allow a region of
stable ferromagnetic fluid.

As explained in Sec.\ \ref{sec:introduction}, it can be arranged that
fermionic molecules confined to a plane experience an additional long-range,
pairwise interaction varying as $\Lambda r^{-3}$, where $r$ is the
interparticle separation and $\Lambda$ is a
constant.\cite{innsbruck,cooper_2009} In a finite, periodic cell, the two-body
interaction between one particle and all the images of another particle that
is at a distance ${\bf r}$ from the first is
\begin{eqnarray} v_A({\bf r}) & = & \sum_{\bf R} \frac{\Lambda}{|{\bf
      r}+{\bf R}|^3} \nonumber \\ & \approx & \sum_{{\bf R} \in S}
  \frac{\Lambda}{|\tilde{{\bf r}}+{\bf R}|^3} + \frac{\Lambda}{A} \int_{R>R_S}
  \frac{d{\bf R}}{|\tilde{{\bf r}}+{\bf R}|^3} \nonumber \\ & \approx &
  \sum_{{\bf R} \in S} \frac{\Lambda}{|\tilde{{\bf r}}+{\bf R}|^3} +
  \frac{2\pi \Lambda}{A R_S} \left(1+\frac{3\tilde{r}^2}{4 R_S^2} \right)
  +O(R_S^{-5}), \nonumber \\  \end{eqnarray} where $S$ is a circular region of
  radius $R_S$ centered on the origin, and $\tilde{\bf r}$ is the minimum
  image of ${\bf r}$.  By making $R_S$ sufficiently large, the approximation
  to the infinite sum can be made arbitrarily good.

The ``Madelung'' term (the interaction of each particle with its own images) is
\begin{eqnarray} v_M & = & \sum_{{\bf R}\neq{\bf 0}} \frac{\Lambda}{|{\bf
      R}|^3} \nonumber \\ & \approx & \sum_{{\bf R} \in S-\{{\bf 0}\}}
      \frac{\Lambda}{|{\bf R}|^3} + \frac{\Lambda}{A} \int_{R>R_S} \frac{d{\bf
      R}}{|{\bf R}|^3} \nonumber \\ & \approx & \sum_{{\bf R} \in S-\{{\bf
      0}\}} \frac{\Lambda}{|{\bf R}|^3} + \frac{2\pi \Lambda}{A R_S}.
      \end{eqnarray}

The Hamiltonian for the finite, periodic cell is therefore
\begin{equation} \hat{H} = -\frac{1}{2} \sum_i \nabla_i^2 +
  \sum_{i>j} \left[ v_H(|\tilde{\bf r}_{ij}|) + v_A(\tilde{\bf r}_{ij})
      \right] + \frac{Nv_M}{2}. \end{equation}

\subsection{Convergence of the real-space sum}

We have chosen $R_S$ such that 119 stars of lattice vectors are summed over
explicitly in square cells (for fluid phases) and 86 stars are summed over in
hexagonal cells (for crystal phases).  The error in the Madelung constant is
about $2\times 10^{-7}$ a.u.\ per particle for $N=18$ particles (and smaller
at larger $N$).  The error in the additional interaction per particle due to
the finite number of stars of ${\bf R}$ vectors should therefore be much
smaller than $10^{-6}$ a.u.  The error made in truncating the real-space sum
is therefore negligibly small compared with the statistical error bars on our
QMC energies.

\subsection{Validity of perturbation theory}

\begin{figure}
\begin{center}
\includegraphics[clip,scale=0.3]{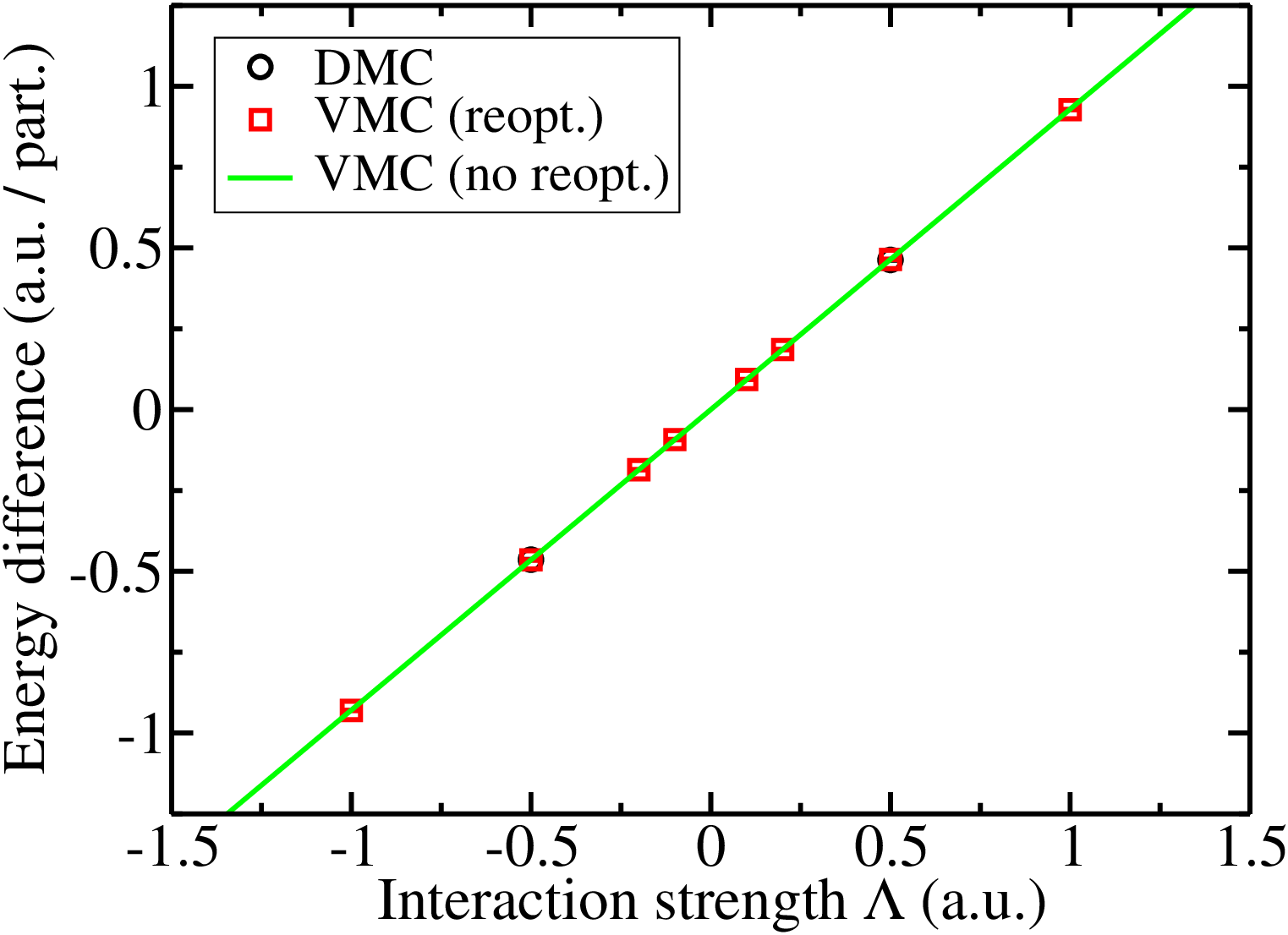}
\caption{(Color online) Change in energy resulting from the inclusion of a
$\Lambda r^{-3}$ interaction for a paramagnetic fluid of $N=18$ hard-core
particles of diameter $D=0.88$ a.u. \label{fig:pert_theory_works}}
\end{center}
\end{figure}

To a very good approximation we can describe the effect of the $\Lambda
r^{-3}$ interaction using perturbation theory.  The change in the energy
resulting from the additional interaction is approximately given by the
expectation of the interaction operator with respect to the previously
optimized wave function for the unperturbed system, which can be evaluated
using VMC\@.  For $-1<\Lambda<1$, this approximation reproduces the full DMC
energy difference to within about 0.002 a.u.\ at $D=0.88$ a.u., as shown in
Fig.\ \ref{fig:pert_theory_works}.

\subsection{Finite-size errors}

\begin{figure}
\begin{center}
\includegraphics[clip,scale=0.3]{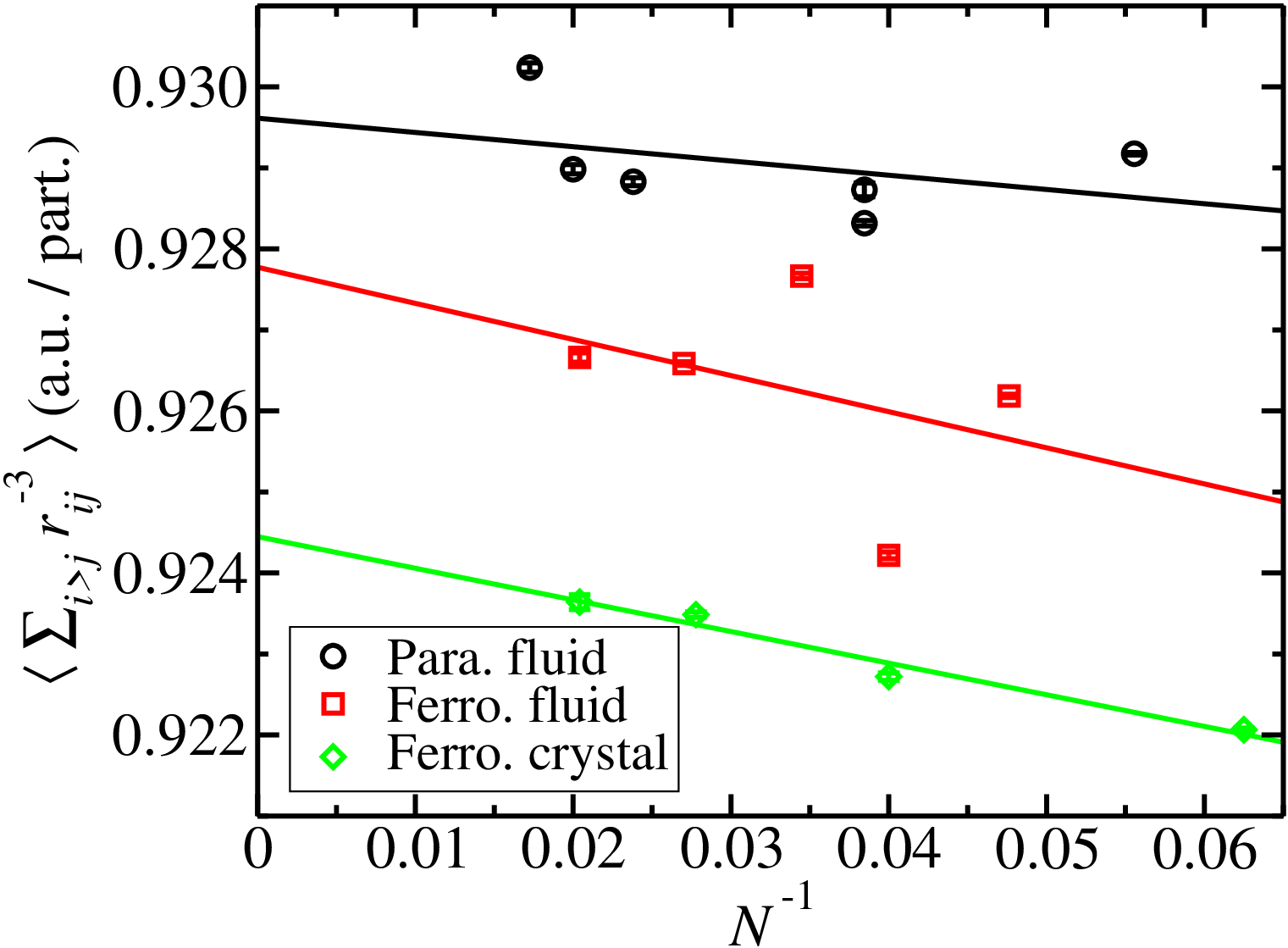}
\caption{(Color online) VMC-evaluated expectation value of $\sum_{i>j}
r_{ij}^{-3}$ with respect to the non-reoptimized VMC wave
function. \label{fig:fs_extrap_of_extra}}
\end{center}
\end{figure}

As with total energies, finite-size errors in the expectation value of the
extra interaction are larger and quasi-random for the unstable phases.  We
extrapolate to infinite system size in each case, as shown in Fig.\
\ref{fig:fs_extrap_of_extra}.

\subsection{Effect of weak interactions on the phase diagram}

\begin{figure}
\begin{center}
\includegraphics[clip,scale=0.3]{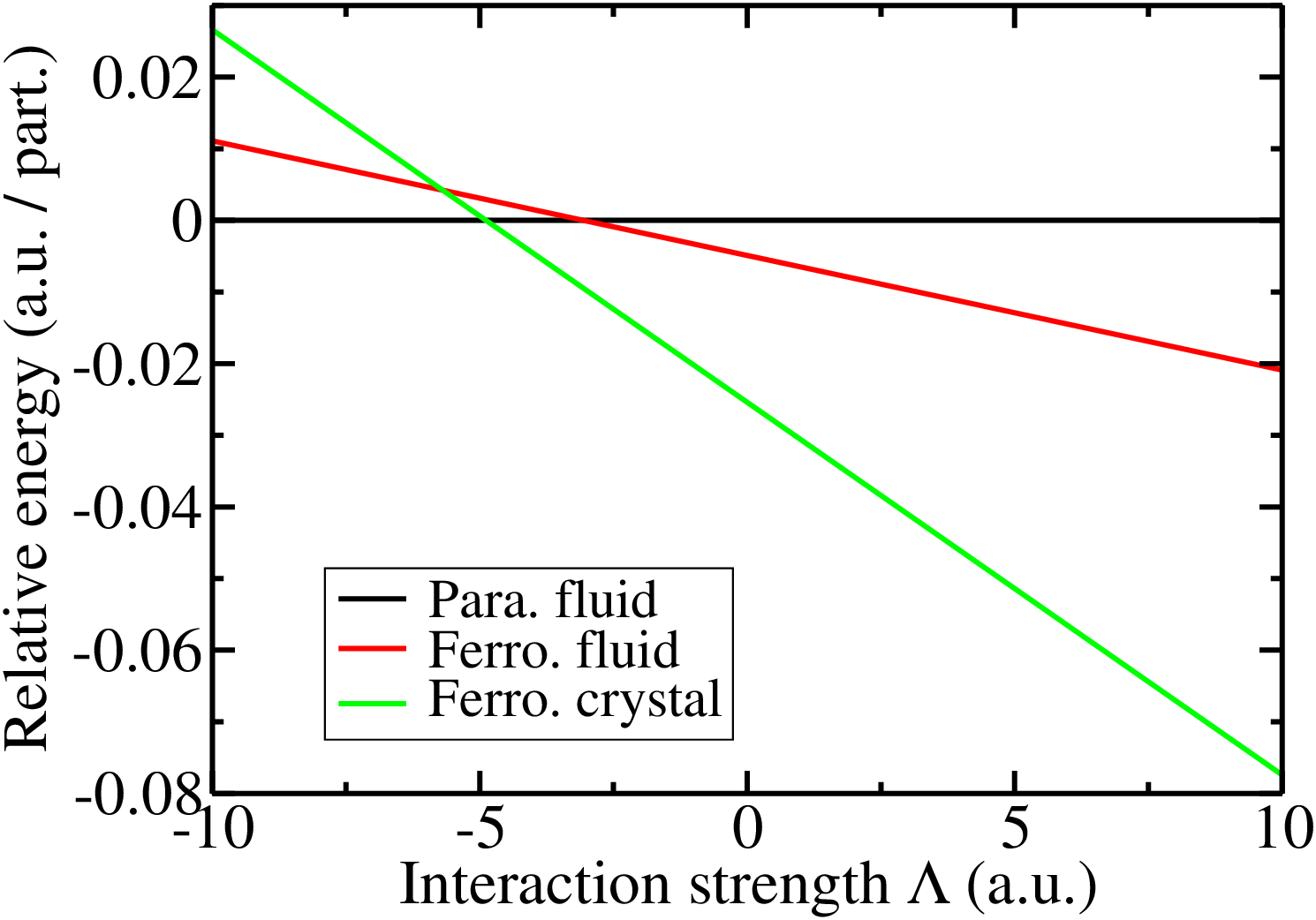}
\caption{(Color online) Energy of the ferromagnetic fluid and crystal relative
  to the energy of the paramagnetic fluid against the strength of the $r^{-3}$
  interaction for hard-core particles of diameter $D=0.88$
  a.u. \label{fig:E_v_Lambda}}
\end{center}
\end{figure}

At $D=0.88$ a.u., combining the VMC results for $\langle \sum_{i>j}
r_{ij}^{-3} \rangle$ with the DMC energies in the absence of the additional
interaction, we find that
\begin{eqnarray} E_{\rm PF} & = & 3.2559+0.9296 \Lambda \nonumber \\
 E_{\rm FF} & = & 3.251+0.928\Lambda \nonumber \\ E_{\rm FC} & = &
 3.2305+0.9244\Lambda, \end{eqnarray} where $E_{\rm PF}$, $E_{\rm FF}$, and
 $E_{\rm FC}$ are the energies in a.u.\ per particle for the paramagnetic
 fluid, ferromagnetic fluid, and ferromagnetic crystals, respectively, as a
 function of $\Lambda$.  The resulting offset in the energy relative to the
 energy of the paramagnetic fluid phase at a density close to the
 crystallization density is shown in Fig.\ \ref{fig:E_v_Lambda}.  The changes
 in the energies of the three phases due to the inclusion of the $\Lambda
 r^{-3}$ tail are very similar.  As one would expect, a repulsive potential
 ($\Lambda>0$) favors phases where the particles are kept apart (i.e.,
 ferromagnetic over paramagnetic phases, and crystals over fluids), whereas an
 attractive interaction ($\Lambda<0$) favors a paramagnetic fluid the most and
 a ferromagnetic crystal the least.  There is no region of stability for the
 ferromagnetic fluid, although it comes close at $\Lambda \approx -5$.
 However, this is a regime where the interaction is strongly attractive, so
 that perturbation theory is no longer valid, and a trial wave function that
 includes the possibility of superfluid pairing is required for accurate QMC
 calculations. It therefore seems unlikely that a weak interaction will have
 much effect on the phase diagram.

\section{Conclusions \label{sec:conclusions}}

Gases of hard-core particles are a natural model for ultracold-atom systems.
Correlation effects in 2D hard-core systems are surprisingly long-ranged.  Our
QMC results show that there is no regime in which itinerant ferromagnetism
occurs in a 2D hard-core fluid.  As the hard-core diameter $D$ is increased,
the system undergoes a transition from a paramagnetic fluid to a crystal at
$D=0.83$ a.u.  The absence of a region of stability for a ferromagnetic fluid
resembles the situation in the 2D HEG\@.\cite{ndd_2d_heg} Including a weak
$r^{-3}$ tail in the two-body interaction between hard-core particles does not
lead to a significant revision of the phase diagram. We have presented QMC
results for the PCFs, static structure factors, and momentum densities of the
fluid phases of the hard-core gas.

\begin{acknowledgments}
We acknowledge financial support from the Leverhulme Trust and the UK
Engineering and Physical Sciences Research Council (EPSRC)\@.  Computing
resources were provided by the Cambridge High Performance Computing Service.
\end{acknowledgments}

\end{document}